\documentclass[a4paper,11pt]{article}
\pdfoutput=1
\usepackage{jheppub}
\usepackage[T1]{fontenc}
\usepackage{multirow}
\usepackage{booktabs}
\usepackage{xcolor}
\usepackage{tabularx}

\usepackage{xspace}
\newcommand{\gev}{\ensuremath{\,\text{GeV}}\xspace}
\newcommand{\tev}{\ensuremath{\,\text{TeV}}\xspace}

\def\be{\begin{equation}}
\def\ee{\end{equation}}
\def\bea{\begin{array}}
\def\beqa{\begin{eqnarray}}
\def\eeqa{\end{eqnarray}}
\def\beqas{\begin{eqnarray*}}
\def\eeqas{\end{eqnarray*}}

\def\bp{\begin{picture}}
\def\ep{\end{picture}}
\def\bc{\begin{center}}
\def\ec{\end{center}}
\def\bfig{\begin{figure}}
\def\efig{\end{figure}}

\def\bit{\begin{itemize}}
\def\eit{\end{itemize}}

\def\f{\frac}

\def\[{\left[}
\def\]{\right]}
\def\({\left(}
\def\){\right)}

\def\..{\left.}
\def\.{\right.}
\def\tl{\tilde}

\def\al{\alpha}

\def\ep{\epsilon}

\title{GUT-scale constrained SUSY in light of new muon g-2 measurement}

\author[a]{Fei Wang,}
\author[b]{Lei Wu,}
\author[c,d]{Yang Xiao,}
\author[c,d]{Jin Min Yang}
\author[a]{and Yang Zhang}

\affiliation[a]{School of Physics and Microelectronics, Zhengzhou University, Zhengzhou 450000, China}
\affiliation[b]{Department of Physics and Institute of Theoretical Physics, Nanjing Normal University,  Nanjing 210023, China}
\affiliation[c]{CAS Key Laboratory of Theoretical Physics, Institute of Theoretical Physics, Chinese Academy of Sciences, Beijing 100190, China}
\affiliation[d]{School of Physics, University of Chinese Academy of Sciences,  Beijing 100049, China }

\emailAdd{feiwang@zzu.edu.cn}
\emailAdd{leiwu@njnu.edu.cn}
\emailAdd{xiaoyang@itp.ac.cn}
\emailAdd{jmyang@itp.ac.cn}
\emailAdd{zhangyangphy@zzu.edu.cn}

\abstract{
The new FNAL result of the muon $g-2$, combined with the BNL result, shows a 4.2$\sigma$ deviation from the SM.
We use the new data of the muon $g-2$ to revisit several GUT-scale constrained SUSY models with the constraints from the LHC searches, the dark matter detection, the flavor data and the electroweak vacuum stability.
We first demonstrate the tension between the muon $g-2$ and other experimental measurements in the CMSSM/mSUGRA.
Then after discussing the possible ways to alleviate such a tension and showing the muon $g-2$ in pMSSM under relevant experimental constraints, 
we survey several extensions of the CMSSM/mSUGRA with different types of universal boundary conditions at the GUT scale. 
Finally,  we briefly discuss the muon $g-2$ in other popular SUSY breaking mechanisms, namely the GMSB and AMSB mechanisms and their extensions. 
}

\begin{document}
\maketitle
\flushbottom

\section{Introduction}

In particle physics the muon anomalous magnetic moment $a_{\mu}=(g-2)_{\mu}/2$ is one of the most precisely measured quantities and its long-standing discrepancy between theory and experiment may be a harbinger for new physics beyond the standard model (SM). In the SM it is predicted as 
\beqa
a_{\mu}^{\rm SM}= (11659181.0\pm4.3) \times 10^{-10},
\eeqa
which comes from theoretical calculations based on $e^+e^-$ data~\cite{Aoyama:2020ynm,Davier:2019can,Davier:2017zfy,Davier:2010nc}.
The previous BNL E821 experimental measurement gave a value  
\beqa
a_{\mu}^{\rm BNL} = (11659209.1 \pm 6.3) \times 10^{-10},
\eeqa
where the experimental error is the quadrature sum of systematic ($3.3\times10^{-10}$) and statistical ($5.4\times10^{-10}$) contributions~\cite{Bennett:2006fi,PDG}. The FNAL E989 experimental just released a new result, which combines with the BNL result gives ~\cite{PhysRevLett.126.141801} 
\beqa
a_{\mu}^{\rm BNL+FNAL} = (11659206.2 \pm 4.1) \times 10^{-10}.
\eeqa
Combining the experimental and theoretical values, the deviation of $a_{\mu}$ is
\beqa
& \Delta a_{\mu}^{\rm BNL} &= (28.1\pm 7.6) \times 10^{-10},\\
& \Delta a_{\mu}^{\rm BNL+FNAL}&= (25.1\pm 5.9) \times 10^{-10}.
\eeqa
So compared with the SM prediction, the old BNL result shows a 3.7$\sigma$ deviation, while  the new combined result of FNAL and BNL shows a 4.2$\sigma$ deviation. Both of them  strongly imply new physics.  

Among new physics theories, the low energy supersymmetry (SUSY) is an overwhelmingly popular candidate, albeit lack of evidence at the LHC. Especially, various Grand Unification~(GUT)-scale constrained SUSY models, which assume different types of universal boundary conditions at the GUT scale, are very well motivated. The most
economical one of such models,  the so-called CMSSM/mSUGRA, has been intensively studied in the past decades\cite{Athron:2017qdc,Han:2016gvr,Ellis:2016tjc,Roszkowski:2014wqa,Buchmueller:2013rsa,Bechtle:2013mda,Henrot-Versille:2013yma,Bornhauser:2013aya,Akula:2013ioa,Citron:2012fg,Buchmueller:2012hv,Strege:2011pk,Bechtle:2012zk,Cao:2011sn,Fowlie:2011mb,Trotta:2008bp,deAustri:2006jwj,Baltz:2004aw,Okada:2016wlm,Tran:2018kxv,Belanger:2017vpq,Fukuyama:2016mqb}. Confronting with the 125 GeV Higgs boson mass and the
muon $g-2$ anomaly, the CMSSM/mSUGRA faces a tension which can be alleviated in various extensions. These extensions usually have different types of universal boundary conditions at the GUT scale, e.g., the model in \cite{Nath:1997qm} has non-universal soft masses in the Higgs sector and in the third generation squark sector, while 
in the model \cite{gSUGRA:WWY} the gluino is much heavier than the electroweak gauginos at the GUT scale. Since the contribution to muon $g-2$ needs light electroweakinos and sleptons while the LHC searches imply heavy colored sparticles and the 125 GeV Higgs mass also require heavy top-squarks, the model in \cite{gSUGRA:WWY} might be favored by solving the tension. On the other hand, due to the large number of free soft parameters at the weak scale, the low energy effective MSSM, also called the phenomenological MSSM~(pMSSM), which does not assume universal boundary conditions at the GUT scale, can readily give sizable contributions to muon $g-2$ under current LHC limits and Higgs boson mass measurement~\cite{Miller:2012opa}. For example, the bino-like dark matter scenario in the phenomenological MSSM, with bino-wino or bino-slepton coanhihlations to give correct dark matter relic density,  can give sizable contributions to muon $g-2$ and the required light sleptons may be covered at the HL-LHC~\cite{Abdughani:2019wai}.      

In this work we focus on the constrained SUSY models, which assume universal boundary conditions at the GUT scale and thus have a much more predictive power than the phenomenological MSSM, and discuss the implication of the FNAL E989 experimental results.
We first demonstrate the tension of the typical CMSSM/mSUGRA, confronting muon $g-2$ with the constraints from the LHC searches.
Then we discuss possible ways to alleviate such a tension and show experimental constraints that can not be alleviated, including the DM relic density, the DM direct detections, the electroweakino and slepton searches, and electroweak vacuum stability. Finally, we survey several extensions of the CMSSM/mSUGRA with different types of universal boundary conditions at the GUT scale, which can satisfy most or all of these experimental constraints.


\section{Muon $g-2$ in CMSSM/mSUGRA}
\label{sec:2}

The CMSSM, motivated by the natural link between SUSY and GUT theories, is a model that employs strong universality conditions at the GUT scale. 
It can be described by only four free parameters plus one sign: the universal scalar mass $M_0$, the universal gaugino mass $M_{1/2}$, the universal trilinear coupling $A_0$, the ratio of the two Higgs vacuum expectation values $\tan\beta$, and the sign of the Higgs/Higgsino mass parameter sgn$(\mu)$.

It is well known that the SUSY contributions to the muon $g-2$ are dominated by the chargino-sneutrino and the neutralino-smuon loops.
 At the leading order of $m_W/m_{\rm SUSY}$ and $\tan\beta$ ($m_{\rm SUSY}$ representing the soft SUSY-breaking masses or the Higgsino mass $\mu$),
 various loop contributions can be approximated as \cite{moroi,Stockinger:2006zn}
 \beqa
 \label{moroi}
 \Delta a_\mu(\tl{W},\tl{H},\tl{\nu}_\mu)&\simeq& 15\times 10^{-9}\left(\f{\tan\beta}{10}\right)\left(\f{(100 {\rm GeV})^2}{\mu~M_2}\right),~  \label{a_mu_loop_1}\\
 \Delta a_\mu(\tl{W},\tl{H},\tl{\mu}_L)&\simeq& -2.5\times 10^{-9}\left(\f{\tan\beta}{10}\right)\left(\f{(100 {\rm GeV})^2}{\mu~ M_2}\right),~  \label{a_mu_loop_2}\\
 \Delta a_\mu(\tl{B},\tl{H},\tl{\mu}_L)&\simeq& 0.76\times 10^{-9}\left(\f{\tan\beta}{10}\right)\left(\f{(100 {\rm GeV})^2}{\mu~ M_1}\right),~ \label{a_mu_loop_3}\\
 \Delta a_\mu(\tl{B},\tl{H},\tl{\mu}_R)&\simeq& -1.5\times 10^{-9}\left(\f{\tan\beta}{10}\right)\left(\f{(100 {\rm GeV})^2}{\mu~M_1}\right),~ \label{a_mu_loop_4}\\
 \Delta a_\mu(\tl{\mu}_L,\tl{\mu}_R,\tl{B})&\simeq& 1.5\times 10^{-9}\left(\f{\tan\beta}{10}\right)\left(\f{(100 {\rm GeV})^2(\mu~M_1)}{m_{\tl{\mu}_L}^2m_{\tl{\mu}_R}^2}\right),~  \label{a_mu_loop_5}
 \eeqa
for loop functions $f_C=1/2$ and $f_N=1/6$. The SUSY contributions to the muon $g-2$~($\Delta a_{\mu}^{\rm SUSY}$) will be enhanced for small soft masses and large $\tan\beta$. It can be seen that large SUSY contributions prefer light sleptons and light electroweakinos. However, current LHC experiments have already set stringent constraints on the masses of colored sparticles, such as the gluino and the first two generations of squarks. The gaugino mass relation is predicted to be $M_1:M_2:M_3\approx 1:2:6$ at the weak scale if the universal gaugino mass assumption at the GUT scale is adopted. With this mass relation, the bino and winos cannot be light since the gluino mass has been pushed heavier than 2.2 TeV. Besides, the universal mass input for sfermions at the GUT scale also sets stringent constraints on the slepton masses at the weak scale. Given the stringent constraints on the squarks, the slepton masses can not be light at the EW scale for a universal sfermion mass input at the GUT scale. The 125 GeV Higgs mass and the vacuum stability also constrain the stop mass and the value of the trilinear Higgs-stop-stop coupling $A_t$, which will relate indirectly to the contribution $\Delta a_\mu$. Thus, it is challenging to explain the muon $g-2$ within the frame work of GUT-scale constrained SUSY, especially the typical CMSSM/mSUGRA. 

\begin{figure}[!th]
\centering 
\includegraphics[width=.49\textwidth]{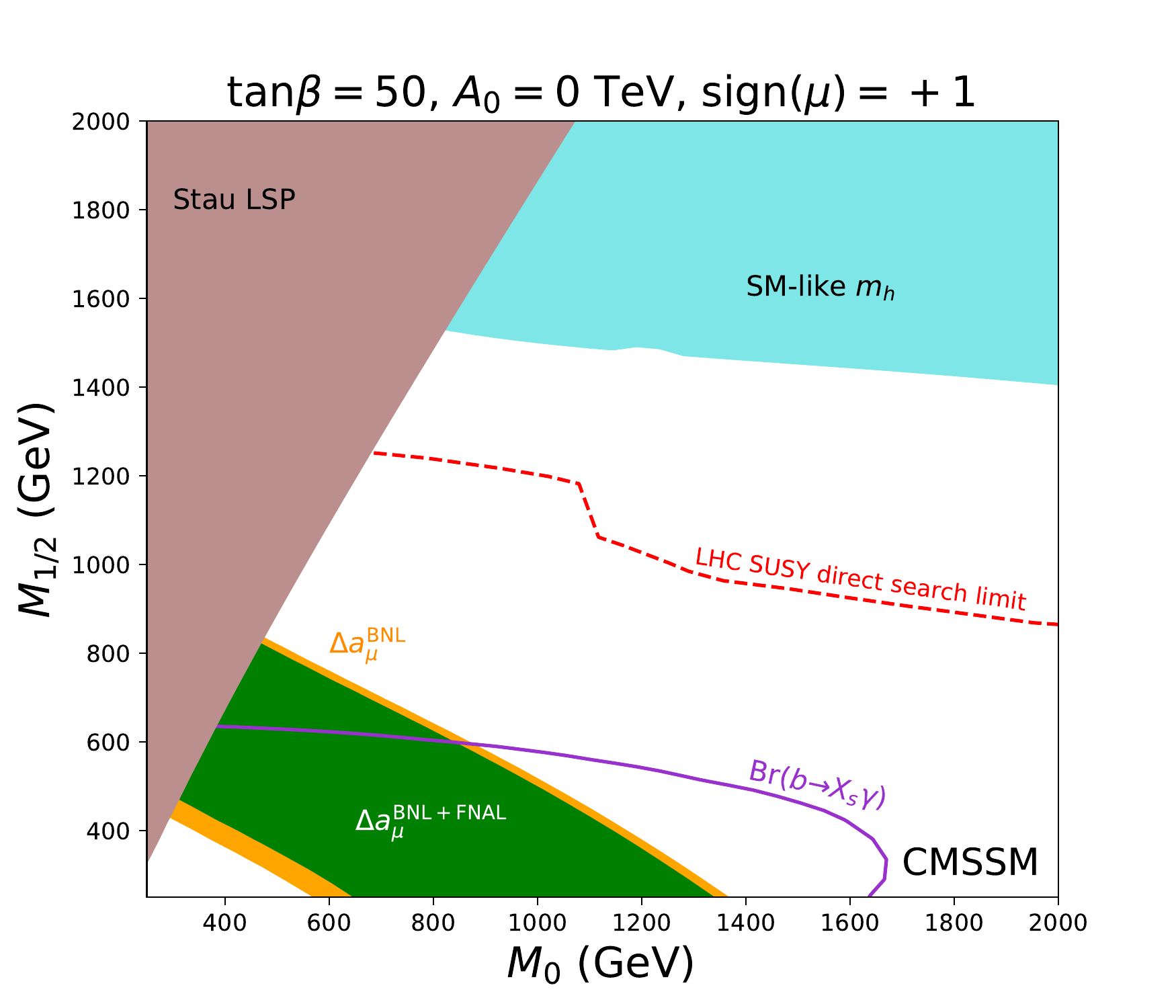}
\includegraphics[width=.49\textwidth]{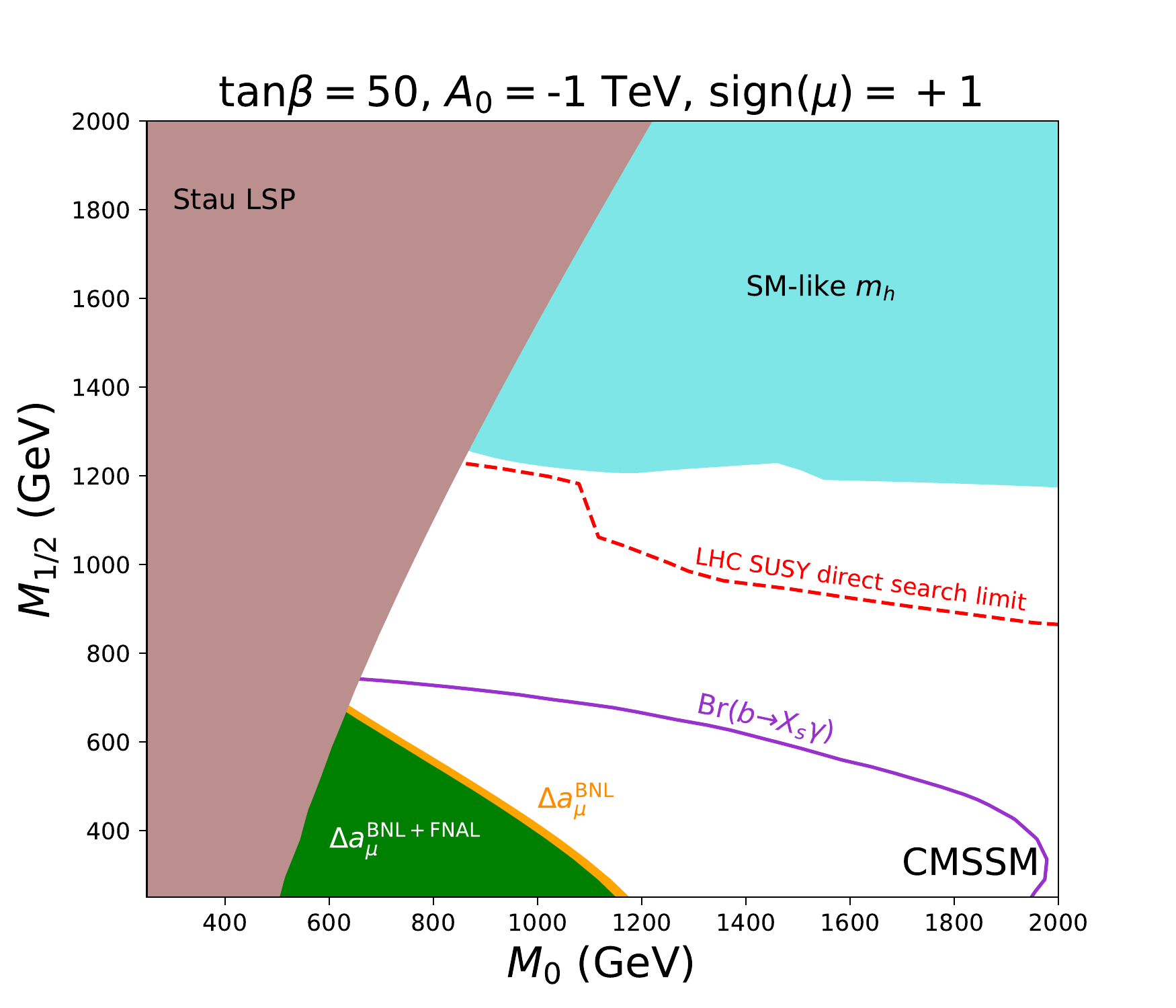}\\
\includegraphics[width=.49\textwidth]{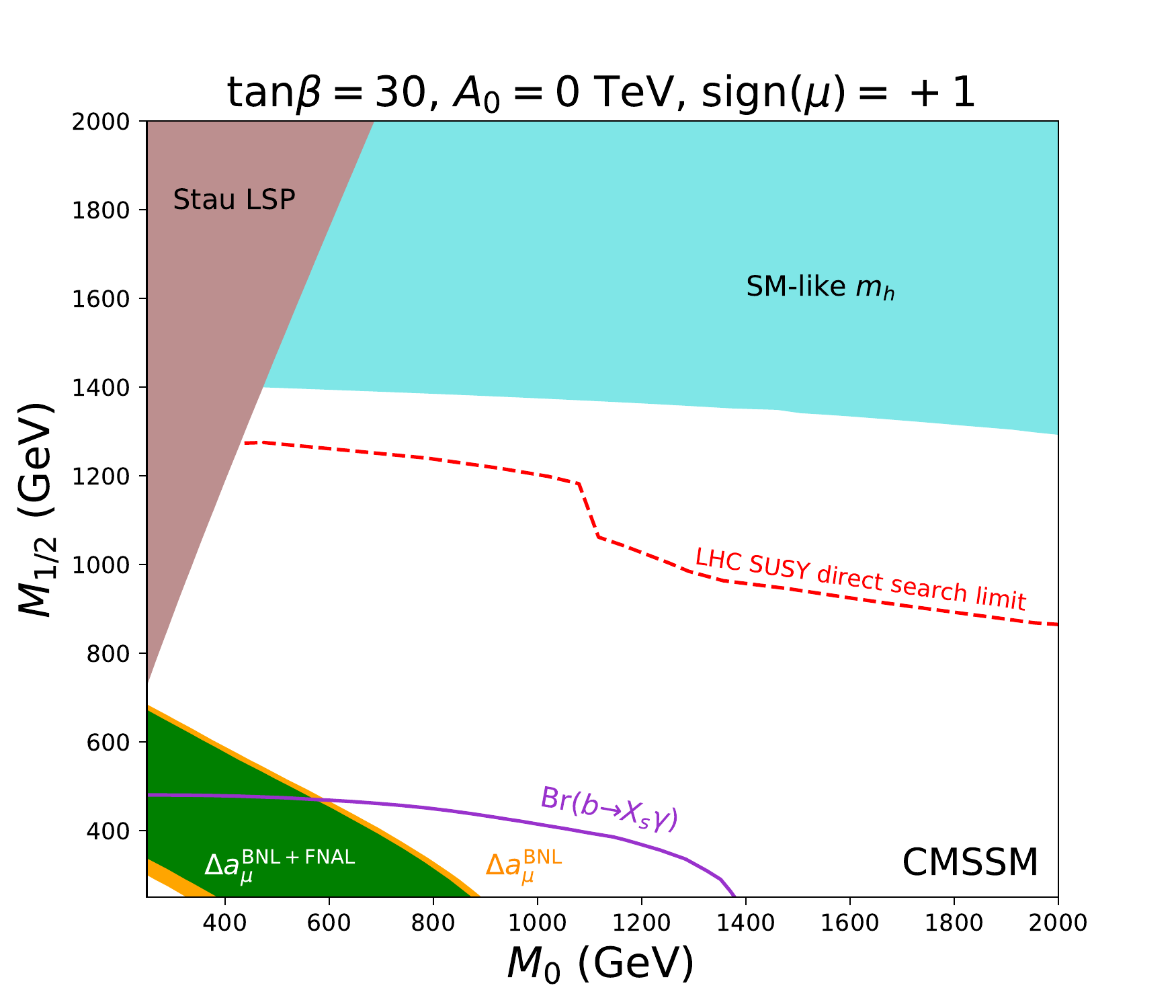}
\includegraphics[width=.49\textwidth]{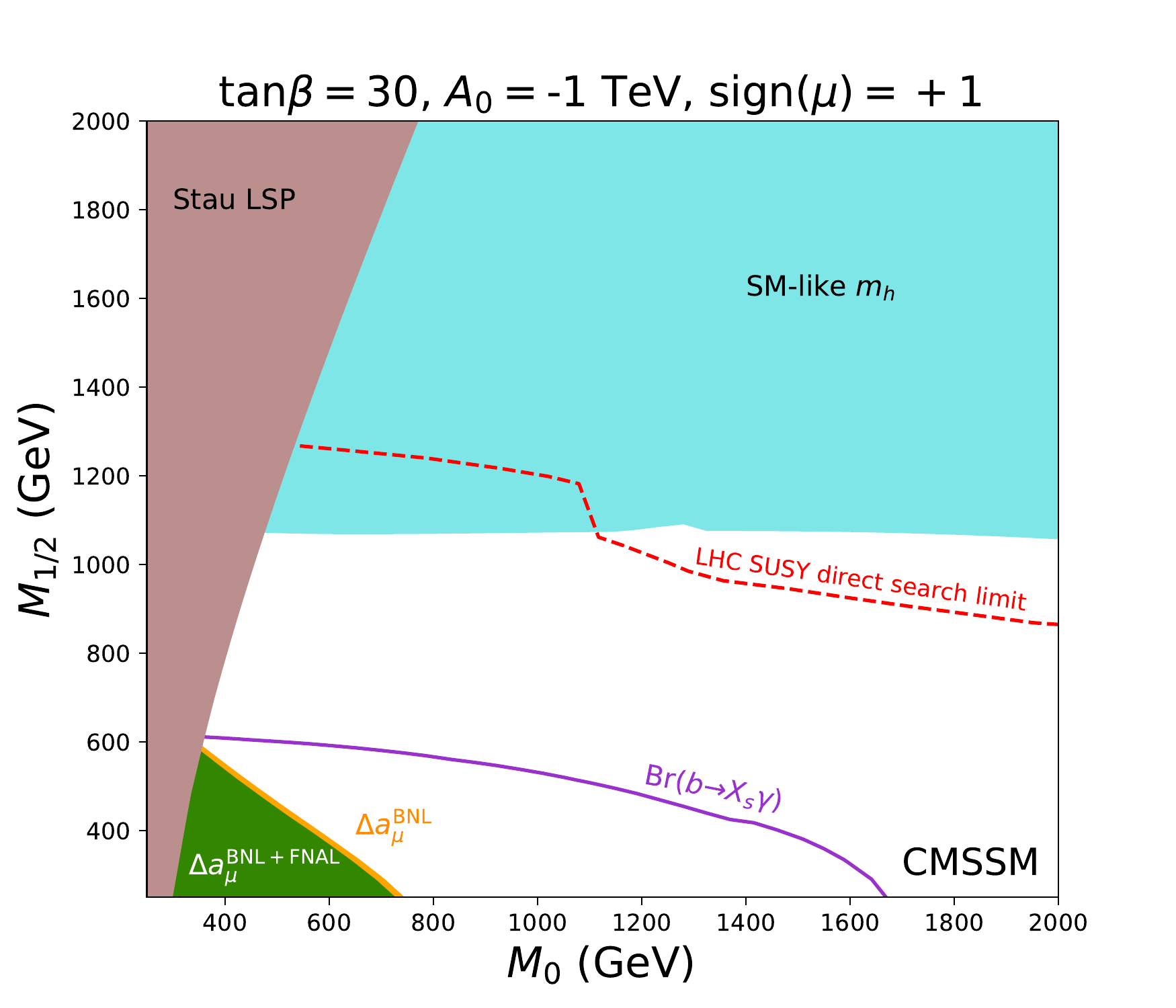}\\
\caption{\label{fig:CMSSM} Tensions between the $\Delta a_{\mu}^{\rm FNAL+BNL/BNL}$ explanation and the relevant experimental limits in CMSSM, showing on the plane of $(M_0,M_{1/2})$ for sign$(\mu)=+1$, $A_0=0$~\tev~(left), 1~\tev~(right) and $\tan\beta=50$~(upper), 30~(lower). The green regions satisfy the combined FNAL+BNL result, while the orange regions plus the green regions are consistent with the previous BNL measurement. The areas below the red dash and purple curves are excluded by LHC direct searches for sparticles and Br($b\to s \gamma$) bound, respectively. The blue shaded regions are consistent with the present Higgs mass measurement, with 2~\gev theoretical uncertainty. In the brown shaded areas, LSP is the charged $\tilde{\tau}_1$, which is excluded. All the limits are shown at $2\sigma$ CL. }
\end{figure}

In Fig.~\ref{fig:CMSSM}, we display the tension between the $\Delta a_{\mu}$ explanation and other experimental limits, using four slices in the parameter space of CMSSM as examples. We fix sign$(\mu)$ = +1, because it is favoured by the measured $\Delta a_{\mu}$~\cite{Feroz:2008wr}.  Here, we use \texttt{SUSY-HIT 1.5}~\cite{Djouadi:2006bz} to generate mass spectrum of sparticles, \texttt{micrOMEGAs 2.4.5}~\cite{Belanger:2001fz,Belanger:2004yn} to calculate DM relic density and direct detection rates, \texttt{SuperIso v3.3}~\cite{Mahmoudi:2008tp} to evaluate flavor physics observables, and \texttt{GM2Calc}~\cite{Athron:2015rva} for calculation of the $\Delta a_\mu$. The interfacing of these programs and the scan is performed by \texttt{EasyScan\_HEP}~\cite{Han:2016gvr}. 

We see from Fig.~\ref{fig:CMSSM} that with the new combined result of the muon $g-2$ the upper limits on $M_0$ and $M_{1/2}$ get stronger. As a result, the conflicts between $\Delta a_{\mu}$ explanation and other experimental limits become more serious, which include
\begin{itemize}
    \item[(i)] LHC direct searches for sparticles. Due to unification of sfermion masses and unification of gaugino masses at the GUT scale, the strongest LHC direct constraints come from the searches for squarks and gluinos in final states of multi-jets and missing transverse momentum~($E_T^{\rm miss}$). The LHC direct search limits at 95\% confidence level~(CL) shown in Fig.~\ref{fig:CMSSM} are estimated by recasting two ATLAS 13 \tev 
    analyses, 2-6 jets + $E_T^{\rm miss}$ (dominate at $M_0<1.3$~\tev)~\cite{ATLAS-CONF-2016-078} and 0/1 lepton + 3 $b$-tagged jets + $E_T^{\rm miss}$(dominate at $M_0>1.3$~\tev)~\cite{ATLAS-CONF-2016-052}, using \texttt{MG5@NLO}~\cite{Alwall:2014hca}, \texttt{PYTHIA}~\cite{Sjostrand:2006za,Sjostrand:2007gs}, \texttt{Delphes3.3.0}~\cite{Ovyn:2009tx,Cacciari:2011ma}, \texttt{Prospino}~\cite{Beenakker:1996ed} and \texttt{CheckMATE 2.0}~\cite{Drees:2013wra,Dercks:2016npn}, as described in Ref.~\cite{Han:2016gvr}.
    The limits, which are insensitive to $\tan\beta$ and $A_0$~\cite{Allanach:2011ut}, have fully excluded parameter space in the CMSSM for explanation of the muon $g-2$ discrepancy. The only way to evade them is to give up or alter the mass unification at the GUT scale.
    \item[(ii)]  The SM-like Higgs measurements. To obtain a $\sim125$~\gev SM-like Higgs boson, a heavy stop or a large $|X_t|=|A_t-\mu\tan\beta|$ is needed to enhance the radiative corrections to $m_h$, which cause lower bounds of about $1(4)~\tev$ on $M_{1/2}(M_0)$, as shown in Fig.~\ref{fig:CMSSM}. In the so-called maximal mixing scenario, where $X_t/\sqrt{m_{\tilde{t}_1}m_{\tilde{t}_2}}\simeq\sqrt{6}$, the contributions from the stops are maximised, and the bounds on $M_{1/2}$ and $M_{0}$ can be relaxed to about 800~\gev and 500~\gev, respectively~\cite{Feng:2013tvd}. Nevertheless, $\tan\beta$ in such a maximal mixing scenario is too small to enhance $\Delta a_{\mu}^{\rm SUSY}$. Performing a global fit of the SM-like Higgs couplings will further increase the conflict.
    \item[(iii)] BR$(B\to X_s\gamma)$ limit. In SUSY models, the SUSY contributions to this decay are dominated by $\tilde{t}\widetilde{W}$ and $tH^{\pm}$ loops. The combination of measurements from CLEO, BaBar, and Belle experiments yields BR$(B\to X_s\gamma)=(3.43\pm0.21\pm0.07)\times10^{-4}$\cite{Amhis:2014hma}, with SM prediction of BR$(B\to X_s\gamma)=(2.98\pm0.26)\times10^{-4}$\cite{Becher:2006pu}. Thus, rather large values of $m_{\tilde{t}}$, $m_{\widetilde{W}}$ and $m_{H^\pm}$ are required to suppress the SUSY loop contributions, which turn into bounds on $M_0$ and $M_{1/2}$, and conflict with the muon $g-2$ explanation. Both the SUSY contributions to BR$(B\to X_s\gamma)$ and $\Delta a_{\mu}^{\rm SUSY}$ increase as $\tan\beta$ increases. Therefore, the conflict
    always exists, roughly independent of  $\tan\beta$. Small values of $|A_0|$ could make some rooms for the muon $g-2$ explanation, but aggravate the tension between $\Delta a_{\mu}^{\rm SUSY}$ and the SM-like $m_h$.
\end{itemize}

It should be noted that the correct relic density is achieved in a delicate way in the parameter space in CMSSM. However, the DM relic density constraint itself does not conflict with the muon $g-2$ explanation. In the bulk region, where neutralino annihilation proceeds dominantly via the t-channel slepton exchange, and the stau co-annihilation region, $M_0$ and $M_{1/2}$ can be small to give rise to the desired $\Delta a_{\mu}^{\rm SUSY}$. We will discuss in detail the relationship between DM relic density and the muon $g-2$ in next section, as well as the DM direct detection limits. 

Therefore, the parameter space in the CMSSM that can yield the required $\Delta a_{\mu}^{\rm SUSY}$ is highly disfavoured by the LHC direct searches for colored sparticles, the SM-like Higgs measurements, and the BR$(B\to X_s\gamma)$ limit, because of the unifications of squarks masses and gaugino masses at the GUT scale. To resurrect such GUT-scale constrained SUSY, we have to alter the unification conditions for the third generation of squarks and gluinos, which will be discussed in Section \ref{sec-4}.

\section{Muon $g-2$ in pMSSM}
\label{sec-3}
Before we discuss the possible ways to relax or alter the unification conditions for the third generation of squarks and gluinos to alleviate the tension suffered by the CMSSM, we check the current experimental limits which are related directly to  $\Delta a_{\mu}^{\rm SUSY}$. Unlike the experimental limits (i-iii) on the masses of the colored sparticles (gluino and stops) which are transferred to the limits on uncolored sparticles (sleptons and electroweakinos) involved in $\Delta a_{\mu}^{\rm SUSY}$ through the assumed boundary conditions at the GUT scale, the direct experimental limits on sleptons and electroweakinos from the LHC searches or DM relic density and DM detections cannot be relaxed by altering the boundary conditions at the GUT scale. We take the pMSSM to demonstrate these robust experimental limits  related directly to $\Delta a_{\mu}^{\rm SUSY}$.

\begin{figure}[t]
\centering 
\includegraphics[width=.49\textwidth]{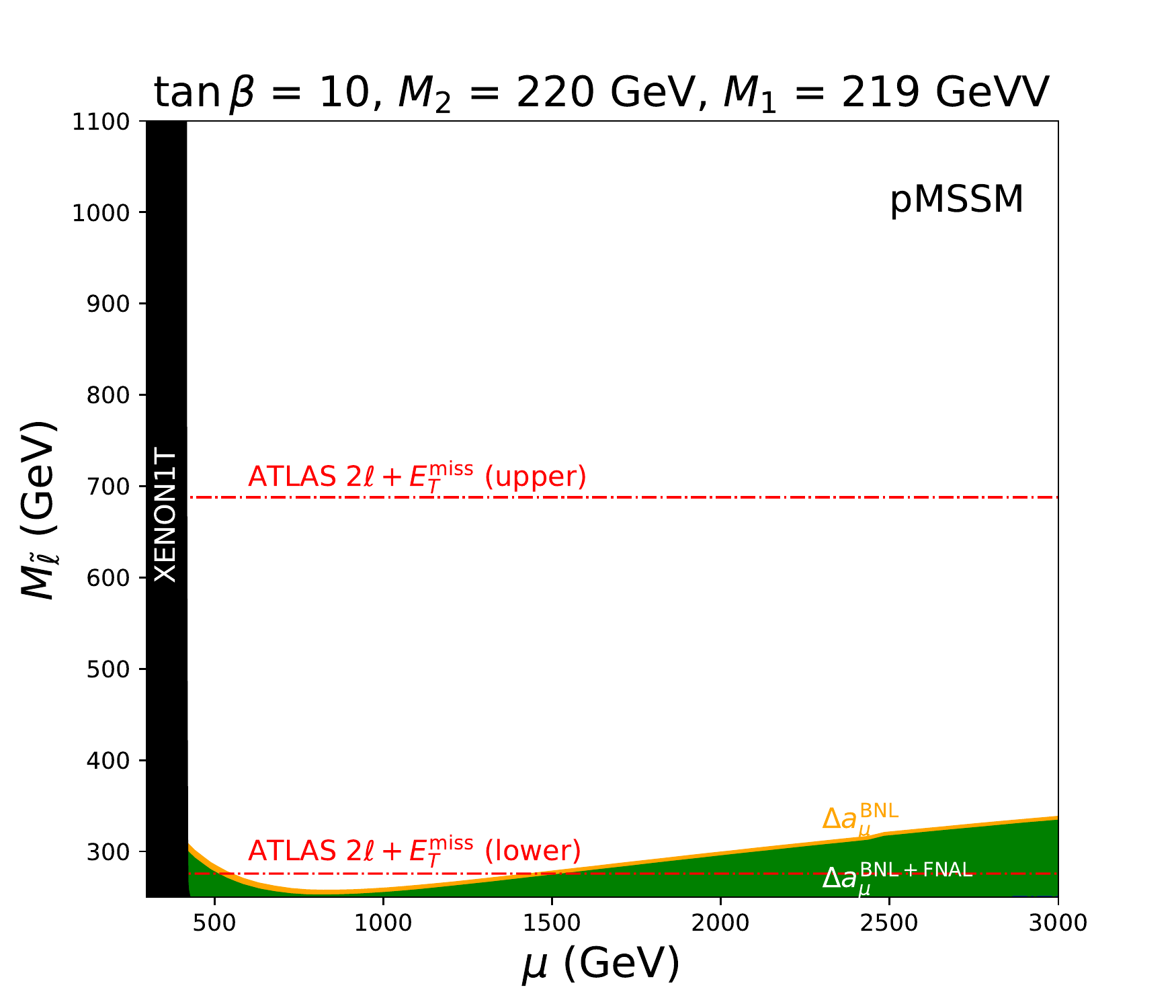}
\includegraphics[width=.49\textwidth]{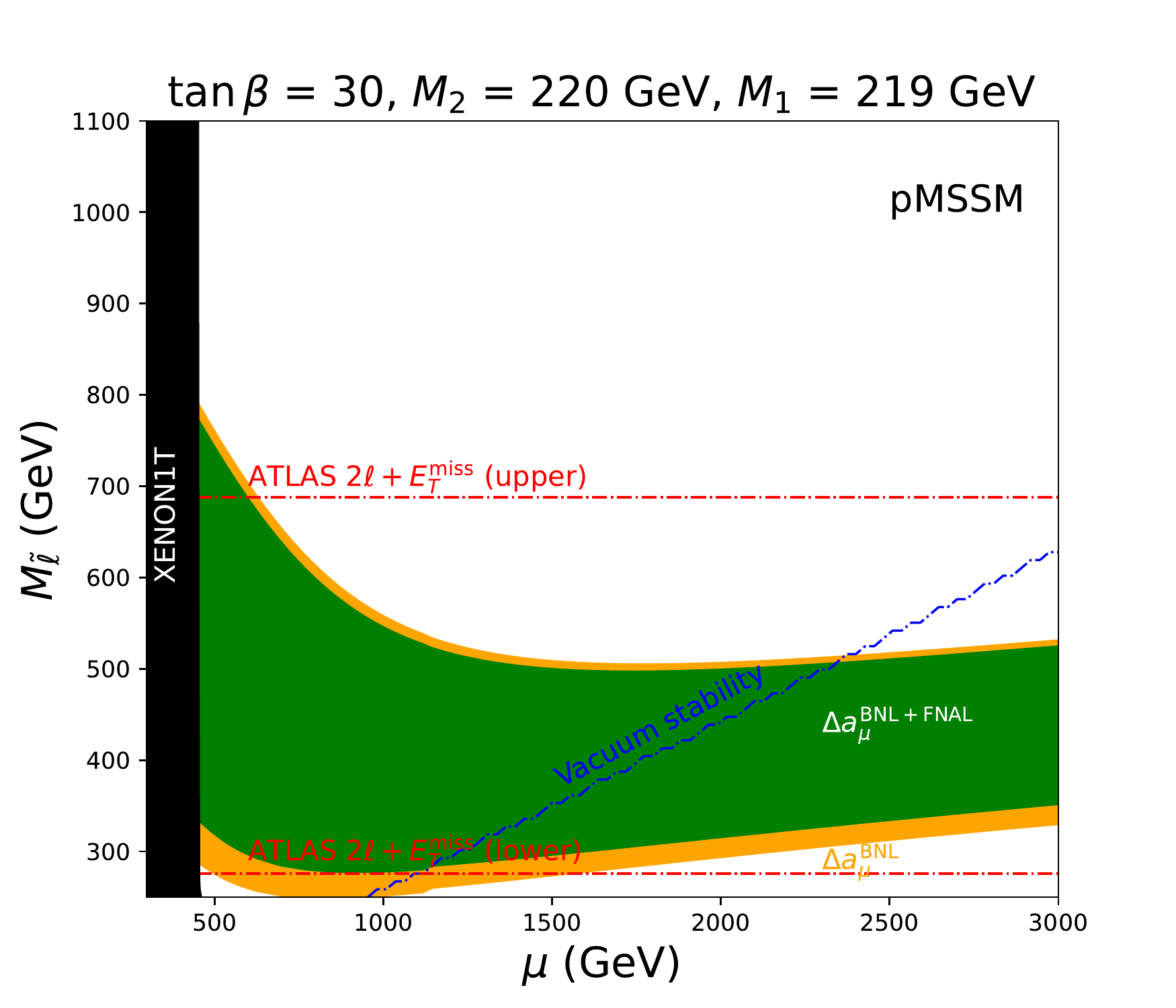}\\
\includegraphics[width=.49\textwidth]{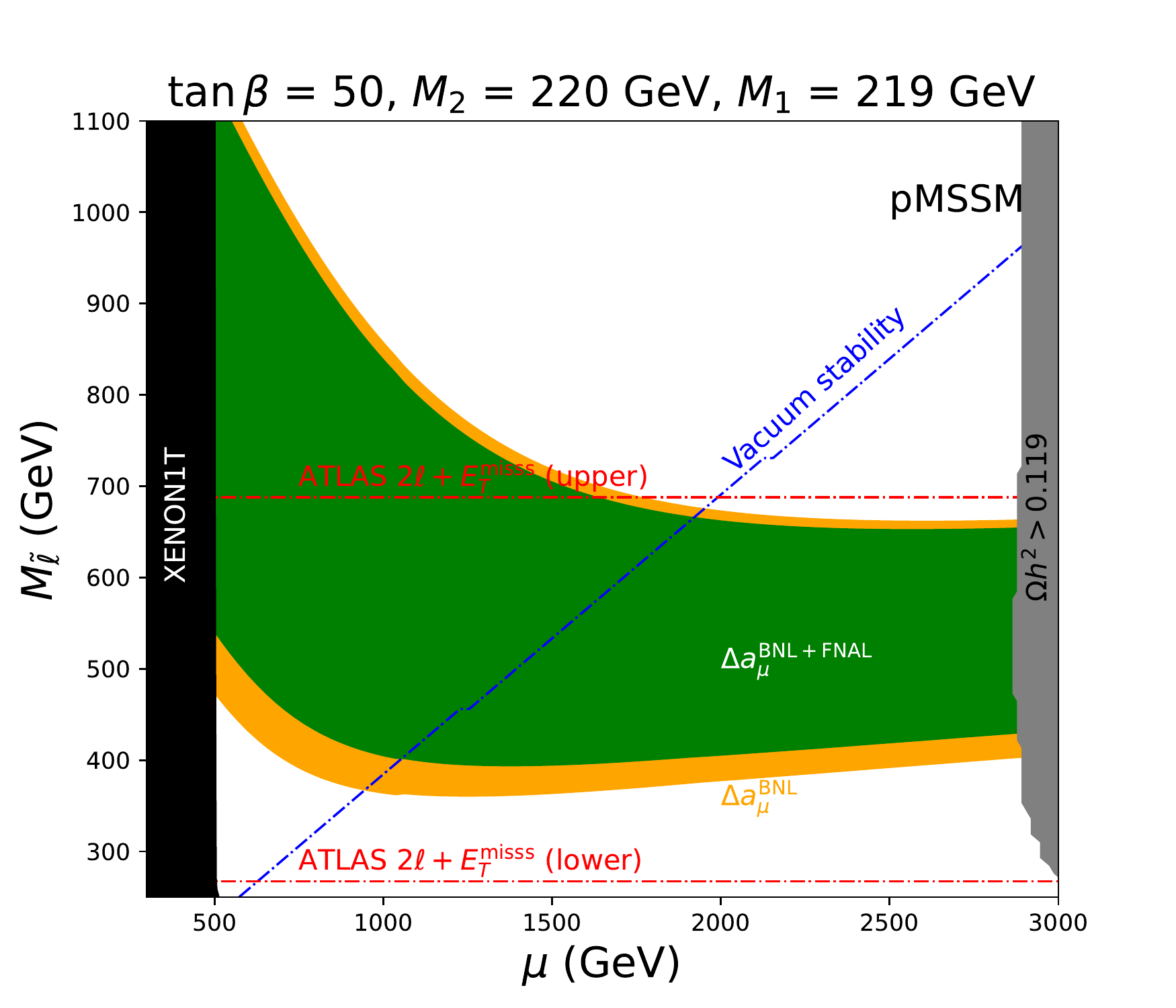}
\includegraphics[width=.49\textwidth]{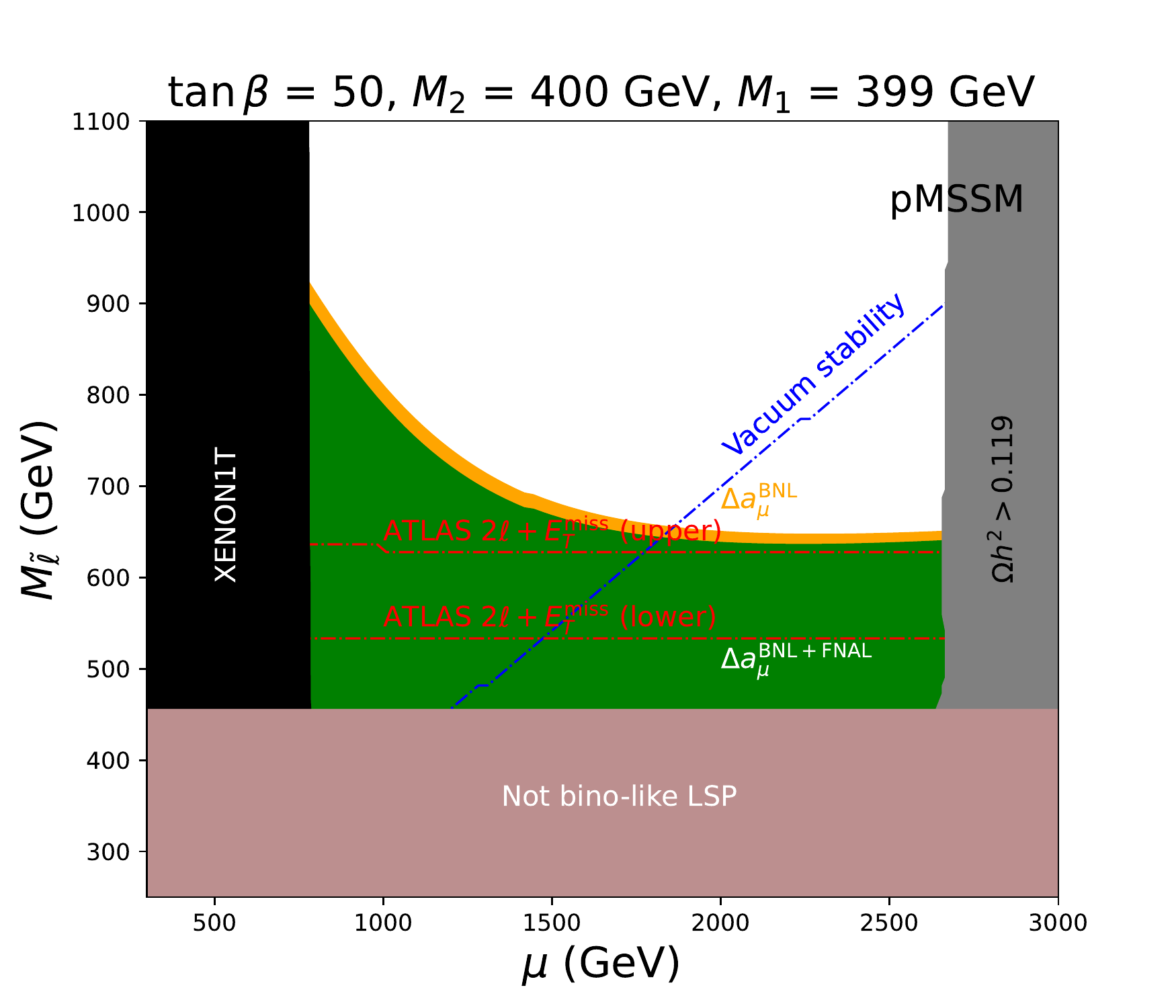}
\caption{\label{fig:BW} FNAL+BNL and BNL $\Delta a_{\mu}$ constraints for the BW scenario in the pMSSM. The orange and dark green regions can explain the BNL and the FNAL+BNL $\Delta a_{\mu}$ measurements at 2$\sigma$ CL. The black region is excluded by Xenon-1T at 90\% CL, while in the brown region the LSP is not bino-like neutralino. The areas between the two ATLAS $2\ell+E_T^{\rm miss}$ limits (red dash lines) are excluded by 13~\tev LHC searches for slepton-pair production at 95\% CL. The regions on the right of blue dash lines spoil stability of the electroweak vacuum.}
\end{figure}

\begin{figure}[t]
\centering 
\includegraphics[width=.49\textwidth]{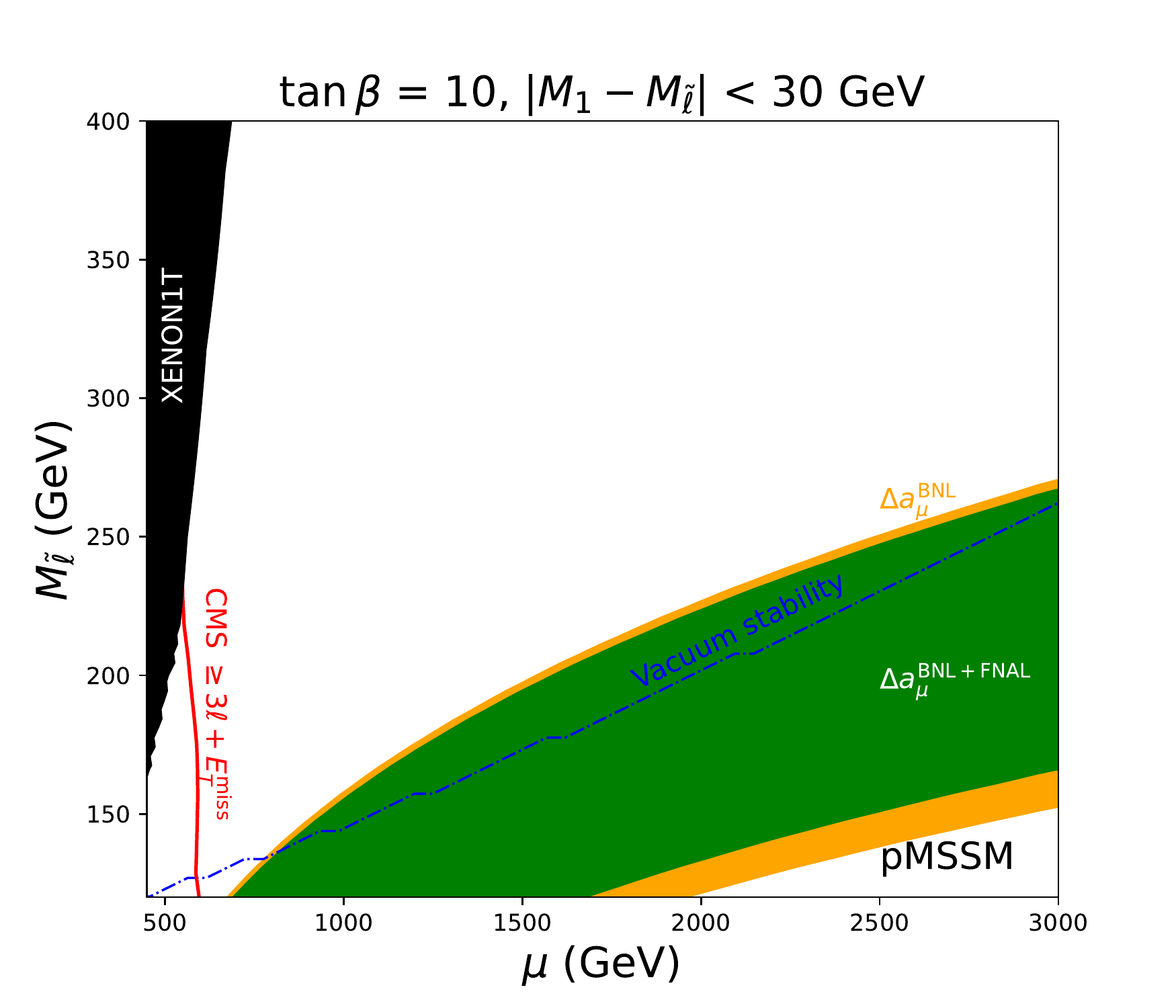}
\includegraphics[width=.49\textwidth]{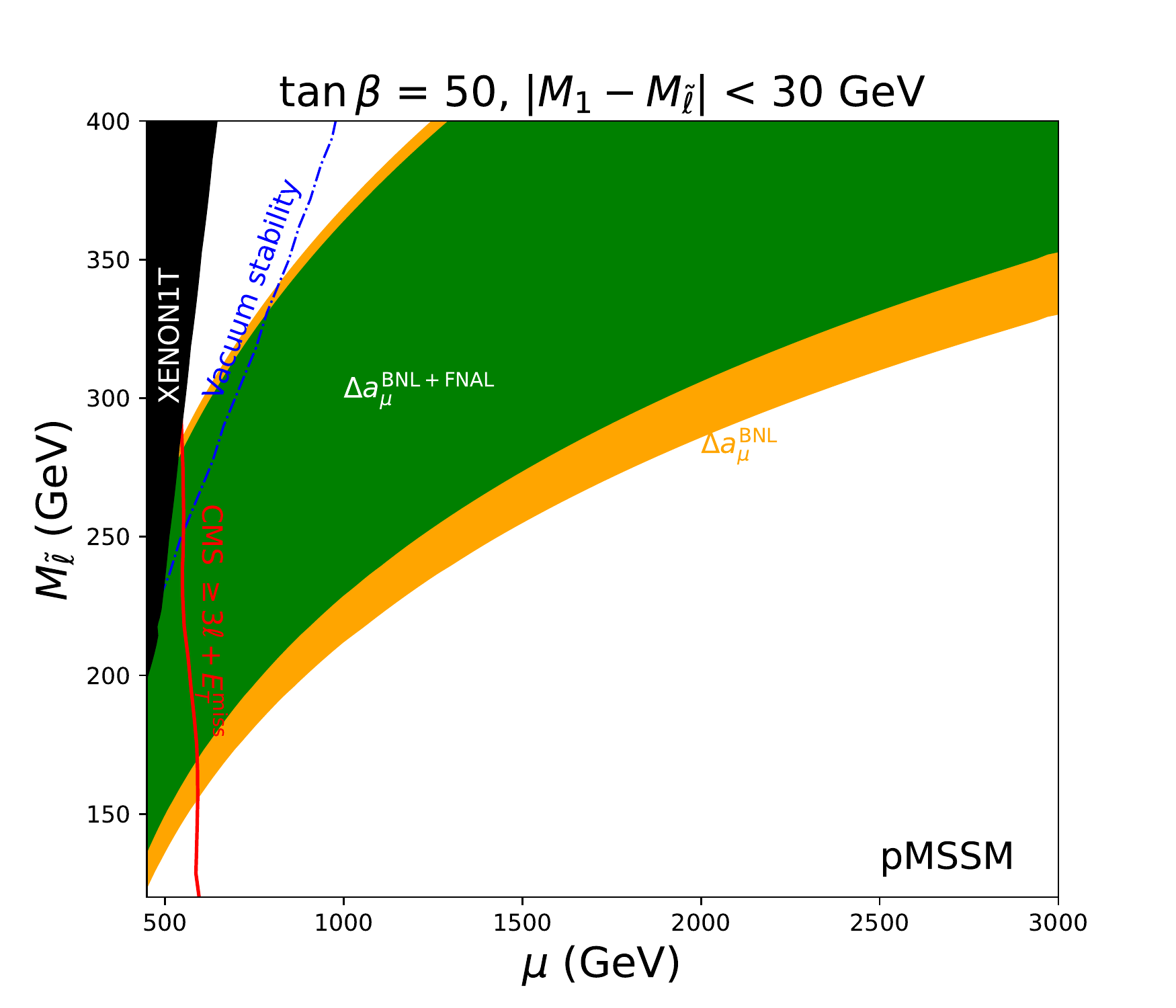}
\caption{\label{fig:BL}Same as Fig.2 but for the BL scenario and the red lines denote the CMS observed exclusion limits at 95\% CL from searches for electroweakinos with $\geq3\ell+E_T^{\rm miss}$ final states.}
\end{figure}

In the pMSSM, we can assume that all coloured sparticles and the additional Higgs bosons are decoupled to meet the null search results at the LHC, and also assume the lightest neutralino is the lightest sparticle (LSP) and plays the role of WIMP DM. We focus on the bino-like dark matter scenarios, in which the bino co-annihilates with wino (BW scenario) or with slepton (BL scenario) to give the correct thermal DM relic density ~\cite{Padley:2015uma,Kobakhidze:2016mdx,Abdughani:2019wai,Cox:2018qyi,Chakraborti:2020vjp,Lindner:2016bgg}.
Numerically, we assume that $M_{\tilde{l}} = M_{\tilde{e}_L} = M_{\tilde{e}_R} = M_{\tilde{\mu}_L} = M_{\tilde{\mu}_R}$, $M_2-M_1=1$~\gev for the BW scenario, $|M_1-M_{\tilde{l}}|<30$~\gev and $M_2=3$~\tev for the BL scenario. The masses of gluino and squarks are fixed at 3~\tev, all the trilinear coefficients are set to zero except that $A_t$ is tuned to obtain a Higgs mass of about 125~\gev, and the CP-odd Higgs mass $M_A$ is pushed to 5~\tev to avoid DM direct detection in some part of the parameter space. $ M_{\tilde{\tau}_L} = M_{\tilde{\tau}_R}$ is also set to 3~\tev to relax LHC constraints
\footnote{When $M_{\tilde{\tau}_L} = M_{\tilde{\tau}_R} = M_{\tilde{l}}$, the desired DM relic density is achieved by neutralino-stau co-annihilation in the BL scenario, which lead to large mass slipping between selectron/smuon and LSP. As a result, the LHC searches for un-compressed slepton can provided strict constraints on slepton masses. See discussion of Tab.\ref{Tab:gluinoSugra} for details of such constraints. }, 
except when we study the electroweak vacuum stability problem. All the parameters are defined at the scale of 3~\tev. 
The regions consistent with the FNAL+BNL and BNL muon $g-2$ measurements at $2\sigma$ CL are displayed in Figs.~\ref{fig:BW} and \ref{fig:BL}. 

In the BW scenario, SUSY contributions to $\Delta a_{\mu}$ described in Eq.~(\ref{a_mu_loop_1}-\ref{a_mu_loop_4}) dominate at small $\mu$, while contribution of Eq.~(\ref{a_mu_loop_5}) starts to take over  when $\mu>1.5$~\tev. As a result, large values of slepton mass are allowed depending on the higgsino mass, as indicated in Fig.~\ref{fig:BW}. With $\tan\beta$ increasing, the preferred region moves towards larger slepton masses rapidly, because all SUSY contributions to $\Delta a_{\mu}$ are proportional to $\tan\beta$. In general, the upper limits on $M_{\tilde{\ell}}$  to interpret the new combined muon $g-2$ deviation are about 1 TeV. Given the mass spectrum required by the BW scenario, this region will be constrained by the following experiments:
\begin{itemize}
    \item LHC direct searches for charginos. With such small wino mass, there are substantial events of wino-like $\tilde{\chi}^{\pm}_1\tilde{\chi}^{0}_2$ pair production at the LHC, but their decay products are extremely soft due to the small mass splitting between $\tilde{\chi}^{\pm}_1/\tilde{\chi}^{0}_2$ and LSP $\tilde{\chi}^{0}_1$. The strongest constraints come from searches for electroweakinos with compressed mass spectra at 13~\tev LHC with 139~fb$^{-1}$ data~\cite{ATLAS-CONF-2019-014}. For the wino-like chargino, $m_{\tilde{\chi}^{\pm}_1}$ of up to 205~\gev for $m_{\tilde{\chi}^{\pm}_1} -m_{\tilde{\chi}^{0}_1}=5~\gev$ are excluded. The $M_2$ values we chose in Fig.~\ref{fig:BW}, $220~\gev$ and $400~\gev$, are roughly safe, but will be covered at the HL-LHC. 
    
    The higgsino mass $\mu$ is also not very heavy in some of the preferred region by the new $\Delta a_{\mu}$ result. The searches for electroweak production of charginos and neutralinos with un-compressed mass spectra have excluded chargino mass up to 1.4~\tev if BR($\tilde{\chi}^{\pm}_1 \to \tilde{\ell} v_{\ell} \to \ell v_{\ell} \tilde{\chi}^{0}_1$)=100\% and BR($\tilde{\chi}^{0}_2\to \tilde{\ell} \ell \to \ell \ell \tilde{\chi}^{0}_1$)=100\%, 600~\gev if BR($\tilde{\chi}^{\pm}_1 \to W \tilde{\chi}^{0}_1$)=100\% and BR($\tilde{\chi}^{0}_2 \to Z \tilde{\chi}^{0}_1$)=100\%, and 250~\gev if BR($\tilde{\chi}^{\pm}_1 \to W \tilde{\chi}^{0}_1$)=100\% and BR($\tilde{\chi}^{0}_2 \to h \tilde{\chi}^{0}_1$)=100\%, assuming $\tilde{\chi}^{\pm}_1$ and $\tilde{\chi}^{0}_2$ are pure winos~\cite{CMS-PAS-SUS-19-012}. In the BW scenario, the higgsino dominated chargino and neutralinos can have all above decay modes, as well as decaying to wino dominated chargino and neutralinos. The dispersed branch ratios make the higgsinos hard to be excluded at the LHC by now.  
    
    \item LHC direct searches for sleptons. As the slepton masses are restricted, the LHC slepton searches~\cite{Aad:2019vnb,CMS-PAS-SUS-20-001} can probe some of the best-fit region for the muon $g-2$. In Fig.~\ref{fig:BW}, the red dash lines indicate the current 95\% CL exclusion limits on a simplified model, where the left- and right-handed selectrons and smuons are assumed to be degenerate, in searches for sleptons decaying into final states with two leptons and $E_T^{\rm miss}$ at 13~\tev LHC with 139~fb$^{-1}$ data ~\cite{Aad:2019vnb}. The regions between upper and lower limits are excluded at 95\% CL. In our scenario, the pole masses of left- and right-handed sleptons are not exactly degenerate, but we have checked that these limits roughly hold, by performing  simulation of a few of points. It is noteworthy that the muon $g-2$ best-fit regions with a modest $\tan\beta$, such as 30, are almost fully excluded by these searches. Regions with larger $\tan\beta$ can escape the constraint with large slepton masses, while the slepton in best-fit regions with smaller $\tan\beta$ are so light that these searches lose sensitivity because of compressed spectrum. A large $M_2$ can also make the spectrum compressed, but is restricted by the DM constraints as discussed later,  so it can not save the modest and small $\tan\beta$ regions. Giving up the assumption of degenerated sleptons, the limits may be weakened about hundreds of \gev, but the regions of a modest $\tan\beta$ are still highly constrained. 
     
    \item DM direct detection. Here we apply the DM relic density constraint as an upper bound only, allowing for the possibility that neutralinos may only be a fraction of DM, and re-scale the neutralino-proton cross-section by the ratio of the predicted relic density to the observed value. Even so, the latest bounds from XENON1T require $\mu>520(430)~\gev$ with $M_2=220~\gev$ and $\tan\beta=50(10)$. For $M_2=400~\gev$, such a bound increases to $\mu>790~\gev$ with $\tan\beta=50$. Meanwhile, the parameter space satisfying the DM relic density constraint shrinks visibly, as well as the best-fit muon $g-2$ region. Consequently, $M_2$ can not be increased too much to escape the LHC searches for sleptons. The LZ experiment can push the limits on $\mu$ to \tev scale, which can exclude the best-fit muon $g-2$ regions that LHC searches can hardly reach. 
    
    \item Electroweak vacuum stability. In the lower panels of Fig.~\ref{fig:BW}, extending range of $\mu$ to very large value, the muon $g-2$ discrepancy may be explained with heavy sleptons that satisfy the LHC searches. However, it will require a large left-right mixing of the sleptons that spoils stability of the electroweak vacuum in the slepton-Higgs potential. We implement such a limit, assuming $M_{\tilde{\tau}_L}=M_{\tilde{\tau}_R}=M_{\tilde{\ell}}$, using the fit formula of the stability condition~\cite{Endo:2013lva}
    \beqa
        |M_{\tilde{\ell}_{LR}}^2| & < \eta_\ell(& 3.06\times10^2 ~\gev \times M_{\tilde{\ell}} - 2.27\times10^4~\gev^2 \\  \nonumber
        & &+1.48\times10^6~\gev^3/M_{\tilde{\ell}} - 2.26\times 10^8~\gev^4/M_{\tilde{\ell}}^2 ~),
    \eeqa
    where $M_{\tilde{\ell}_{LR}}^2$ is the off-diagonal component of the slepton mass matrix, 
    \beqa
        M_{\tilde{\ell}_{LR}}^2 = -\frac{m_{\ell}}{1+\Delta_\ell} \mu \tan\beta,
    \eeqa
    and $\Delta_\ell$ is a correction to the lepton Yukawa coupling. In the large $\tan\beta$ region, it is given as 
    \beqa
        \Delta_\ell = -\mu\tan\beta \[ \frac{3}{32\pi^2} g^2 M_2 I(M_{\tilde{\ell}}^2, M_2^2, \mu^2) - \frac{1}{16\pi^2} g'^2 M_1 I(M_{\tilde{\ell}}^2,M_{\tilde{\ell}}^2,M_1^2) \],
    \eeqa
    where the loop function is defined as 
    \beqa
        I(a,b,c) = -\frac{ab\ln{(a/b)}+bc\ln{(b/c)}+ca\ln{(c/a)}}{(a-b)(b-c)(c-a)}.
    \eeqa
    In the limit of $a=b$, it becomes
    \beqa
        I(a,a,c) = -\frac{(c-a)+c\ln(a/c)}{(c-a)^2}.
    \eeqa
    We take numerical value of $\eta_\tau$ from \cite{Kitahara:2013lfa}, i.e. 0.9 for $\tan\beta=10$, 0.92 for $\tan\beta=30$ and 0.95 for $\tan\beta=50$.

    In Fig.\ref{fig:BW}, the best-fit muon $g-2$ regions excluded by this limit have already been disfavoured by the LHC slepton searches, but it forbids the possibility of extending $\mu$ to very large values.
\end{itemize}

In the BL scenario, the $\tilde{\mu}_L$-$\tilde{\mu}_R$-$\widetilde{B}$ loop contribution in Eq.~\ref{a_mu_loop_5} dominates $\Delta a_{\mu}^{\rm SUSY}$. As a result, in the best-fit muon $g-2$ regions, $M_{\tilde{\ell}}$ increases as $\mu$ increases, whose upper limits are lower than in the BW scenario. The relevant experimental constraints are similar to the BW scenario, but with different behaviours:
\begin{itemize}
    \item LHC direct searches for sleptons. Similar to the wino in the BW scenario, the sleptons in the BL scenario are very light and have small mass splittings with LSP. Assuming mass-degenerated selectrons and smuons, the slepton masses below 256~\gev are excluded for $m_{\tilde{\ell}} -m_{\tilde{\chi}^{\pm}_1}=10~\gev$ by the ATLAS searches~\cite{ATLAS-CONF-2019-014}, but the limits drop sharply with smaller mass splittings. In the parameter space shown in Fig.\ref{fig:BL}, the splittings between the masses of slepton and bino-like LSP are too small to be detected at the LHC.
    
    \item LHC direct searches for charginos. As the higgsino-like $\tilde{\chi}_1^{\pm}$ mainly decay to $W\tilde{\chi}^0_1$ and BR($\tilde{\chi}_2^0\to Z \tilde{\chi}_1^0$)+BR($\tilde{\chi}_3^0\to Z \tilde{\chi}_1^0$) $\simeq$ 100\%,  independent of $\tan\beta$~\cite{Calibbi:2014lga}, they can be probed at the LHC by searching for events of multi-leptons plus $E_T^{\rm miss}$. The CMS limits displayed in Fig.~\ref{fig:BL} are obtained from searching for events with three or more leptons with 13 TeV 137 fb$^{-1}$ data, assuming that the wino-like $\tilde{\chi}_1^{\pm}\tilde{\chi}_2^0$ pair decaying as $\tilde{\chi}_1^{\pm}\tilde{\chi}_2^0\to ZW\tilde{\chi}_1^0\tilde{\chi}_1^0$~\cite{CMS-PAS-SUS-19-012}.  Although the pair production of winos is larger than pair production of the higgsinos, the limits we used are not overestimated, because the higgsino-like neutralino that decays to $h \tilde{\chi}_1^0$ can provide also some signal events ~\cite{CMS-PAS-SUS-20-003}, and the combination of $\leq 3\ell+E_T^{\rm miss}$ searches and $2\ell +E_T^{\rm miss}$ searches can visibly strengthen the limits~\cite{Pozzo:2018anw}.
    
    \item DM direct detection. We adjust $M_1$ around $M_{\tilde{\ell}}$ to acquire the correct DM relic density. This requires the mass splitting between LSP and lightest slepton shrink quickly as $M_{\tilde{\ell}}$ increases. It is hard to get the required DM relic density when $M_{\tilde{\ell}}>400~\gev$. The XENON1T limits here are similar to the BW case, excluding a small $\mu$ because of the higgsino component in LSP. Due to the same reason, the constraints become slightly stronger with heavier $M_1$.
    
    \item Electroweak vacuum stability. With lower slepton masses required by the desired $\delta a_{\mu}$, the limits of vacuum stability in the BW scenario are much more important than the BL case. For $\tan\beta=50$, there is a tiny region that can satisfy the new muon $g-2$ constraint, the LHC searches for electroweakinos and the electroweak vacuum stability simultaneously. The constraints will get loose if $M_{\tilde{\tau}}\gg M_{\tilde{\ell}}$. However, the assumption of unified slepton masses are very common in GUT-scale constrained SUSY models, and non-universal slepton masses may lead to lepton flavour violation, and CP violation~\cite{Cox:2018qyi,Endo:2013lva}.
\end{itemize}

Generally speaking, the explanation of the new muon $g-2$ measurement in both the BW and BL scenarios are tightly constrained by the latest LHC Run-2 results from the  searches for electroweakinos and sleptons, the DM relic density and electroweak vacuum stability, especially the region with a modest $\tan\beta$ in the BW scenario and the regions with a large $\tan\beta$ in the BL scenario. However, there are still several feasible parameter spaces in the pMSSM for muon $g-2$ and dark matter experiments, which offer opportunity for extending GUT-scale constrained SUSY, and are able to be detected at future hadron colliders, such as the HL-LHC,
HE-LHC and FCC-hh/SPPC~\cite{Kobakhidze:2016mdx,Baum:2020gjj}. 

\section{Muon $g-2$ in CMSSM/mSUGRA  extensions}
\label{sec-4}

Now we discuss various extensions of the CMSSM/mSUGRA-type which can possibly 
solve the muon $g-2$ anomaly:
\begin{itemize}
\item
The first possibility is to relax the gaugino mass ratio~\cite{Ellis:1985jn,Drees:1985bx,Gogoladze:2014cha,Jung:2013zya,Younkin:2012ui,Martin:2013aha,Chakrabortty:2008zk,Martin:2009ad,Bhattacharya:2007dr,Ananthanarayan:2007fj,Chakraborti:2014fha,Wang:2015mea,Balazs:2010ha}. In the non-universal gaugino mass scenarios, the gaugino mass inputs at the GUT scale are not universal. For example, with ${\bf 24}$, ${\bf 75}$ and ${\bf 200}$ representation Higgs fields of SU(5) GUT group, the gaugino mass ratio at the GUT scale can possibly be $M_1:M_2:M_3\approx 1:3:-2$, $M_1:M_2:M_3\approx -5:3:1$ or
$M_1:M_2:M_3\approx 10:2:1$. This amounts to the approximate gaugino ratio $M_1:M_2:M_3\approx 1:6:-12$ for ${\bf 24}$, $M_1:M_2:M_3\approx -5:6: 6$ for ${\bf 75}$ and $M_1:M_2:M_3\approx 5:2:3$ for ${\bf 200}$ at the EW scale, respectively. 
So, only the choice {\bf 24} representation Higgs can lead to lighter bino for 2.2 TeV gluino. This scenario can marginally solve the muon $g-2$ anomaly in the $3\sigma$ range.
SO(10) and $E_6$ GUT can also lead to various possibility of gaugino ratios, depending on the choices of GUT breaking. In general, large ratios between $M_3$ and $M_1,M_2$ are always welcome to solve the muon $g-2$ anomaly.

\item 
An especially interesting scenario is the gluino-SUGRA ($\tl{g}$SUGRA) scenario proposed in \cite{Akula:2013ioa}. The gluino mass can be much heavier than the other gauginos and sfermions at the unification scale. Therefore, at the weak scale the electroweakino masses are not correlated to gluino mass. Besides, through the renormalization group equations (RGE) the squark masses will be driven to values proportional to the gluino mass as they run down from the GUT scale to the weak scale, ameliorating the stringent constraints from the LHC squark searches. The sleptons, which carry no color charge, will stay light. So, the RGE evolution will split the squarks masses and slepton masses at the weak scale for a common $m_0$ at the GUT scale. As shown in \cite{Akula:2013ioa}, a much heavier gluino can be realized by certain combination of {\bf 75} and {\bf 24} representation Higgs. Two of us also proposed to generate a much heavier $M_3$ parameter via a properly chosen gauge kinetic terms~\cite{gSUGRA:WWY,gSUGRA:WWYZ}. 

\begin{table}[th]
\centering
\small
\begin{tabularx}{\textwidth}{Xcccccccc}
\toprule
\multicolumn{7}{c}{Input parameters at GUT scale}                                                                         \\
\midrule
            & $M_{0}$		& $M_{3}$	& $M_{1/2}$ 	& $A_0$ 	& $\tan\beta$	& sign$(\mu)$                     \\
\texttt{P1} & 243.4~\gev	& 3054.3~\gev & 421.6~\gev 	&  120.8~\gev     & 6.1              & +1                                  \\
\texttt{P2} & 520.1~\gev	& 3315.5~\gev & 207.8~\gev 	&  264.2~\gev     & 39.3              & +1                                  \\
\bottomrule
\toprule
\multicolumn{7}{c}{Parameters at SUSY scale}                                                                         \\
\midrule
            & $M_{\rm SUSY}$ & $M_1$       & $M_2$      & $M_3$ & $M_{\tilde{\mu}_L}$ & $M_{\tilde{\mu}_R}$  \\
\texttt{P1} &1000~\gev         	   & 161.9~\gev  & 275.5~\gev &  6402.2~\gev  &  267.5~\gev &    265.3~\gev         \\   
\texttt{P2} &1000~\gev         	   & 438.1~\gev  & 97.6~\gev &  6943.2~\gev  &  501.4~\gev &    630.4~\gev         \\   
\midrule
            & $M_{\tilde{\tau}_L}$ & $M_{\tilde{\tau}_R}$  	& $M_{\tilde{Q}_{1,2,L}}$  & $M_{\tilde{Q}_{3,L}}$ &	$M_{\tilde{u}_R}$ & $M_{\tilde{t}_R}$ \\
\texttt{P1} & 267.5~\gev            &     264.6~\gev 		& 5432.8~\gev 		& 5082.5~\gev 	&   5451.3~\gev		& 4723.7~\gev 	\\
\texttt{P2} & 521.4~\gev            &     662.5~\gev 		& 5896.7~\gev 		& 5384.5~\gev 	&   5924.7~\gev		& 5164.6~\gev 	\\
\bottomrule
\toprule
\multicolumn{7}{c}{Mass spectrum}                                                                         \\
\midrule
            & $m_{h}$ 		& $m_{H}$  	 & $m_{\tilde{\chi}^0_1}$  & $m_{\tilde{\chi}^0_2}$ & $m_{\tilde{\chi}^0_3}$ & $m_{\tilde{\chi}^{\pm}_1}$ \\
\texttt{P1} & 123.6~\gev       & 3363.5~\gev  & 161.0~\gev 			 & 287.7~\gev 			&   3327.2~\gev		& 294.7~\gev 	\\
\texttt{P2} & 123.6~\gev       & 2795.2~\gev  & 108.5~\gev 			 & 441.3~\gev 			&   3464.4~\gev		& 108.9~\gev 	\\
\midrule
            & $m_{\tilde{e}_L}$ & $m_{\tilde{e}_R}$ & $m_{\tilde{\nu}_{e}}$  & $m_{\tilde{\tau}_1}$  & $m_{\tilde{\tau}_2}$ & $m_{\tilde{\nu}_{\tau}}$\\
\texttt{P1} & 203.9~\gev       & 332.5~\gev 		  & 255.7~\gev  & 162.6~\gev 			 & 349.7~\gev 		 	& 252.6~\gev \\
\texttt{P2} & 383.7~\gev       & 683.6~\gev 		  & 535.1~\gev  & 305.4~\gev 			 & 829.3~\gev 		 	& 604.6~\gev \\
\midrule
            & $m_{\tilde{g}}$  & $m_{\tilde{t}_1}$ & $m_{\tilde{t}_2}$  & $m_{\tilde{b}_1}$  & $m_{\tilde{u}_L}$ & $m_{\tilde{u}_R}$\\
\texttt{P1} & 6192.6~\gev      & 4492.1~\gev      & 4846.3~\gev  & 4831.3~\gev 		     & 5203.2~\gev 		 	& 5214.3~\gev \\
\texttt{P2} & 6697.0~\gev      & 4901.1~\gev      & 5146.8~\gev  & 5112.7~\gev 		     & 5624.2~\gev 		 	& 5650.8~\gev \\
\bottomrule
\end{tabularx}
\begin{tabularx}{\textwidth}{lcccccccc}
\toprule
\multicolumn{5}{c}{Branch ratio}    \\
\midrule
            & BR$(\tilde{\chi}_1^{\pm}\to\tilde{e}_L^{\pm}\nu_e)$ & BR$(\tilde{\chi}_1^{\pm}\to\tilde{\chi}_1^0 W^{\pm})$  & BR$(\tilde{\chi}_2^{0}\to\tilde{e}_L^{\pm}e^{\mp})$  &  BR$(\tilde{\chi}_2^{0}\to\tilde{\chi}_1^0 Z)$  \\
\texttt{P1}~~~~~~~~ & 27.9\%  & $<1\%$     & 28.7\%  &  $<1\%$ \\
\texttt{P2}~~~~~~~~ & 0  & 0     & 4.5\%  &  $<1\%$ \\
\midrule
            & BR$(\tilde{e}_L^{\pm}\to\tilde{\chi}_1^0e^{\pm})$ & BR$(\tilde{e}_L^{\pm}\to\tilde{\chi}_1^{\pm} \nu_{e})$  & BR$(\tilde{e}_R^{\pm}\to\tilde{\chi}_1^0e^{\pm})$  &   BR$(\tilde{e}_R^{\pm}\to\tilde{\chi}_2^0e^{\pm})$ \\
\texttt{P1} & 100\%  & 0     & $\sim 100$\%  &  $<1\%$ \\
\texttt{P2} & 33.47\%  & 66.6\%     & $<1\%$  &  $\sim 100\%$ \\
\bottomrule
\end{tabularx}

\begin{tabularx}{\textwidth}{lcccccc}
\toprule
\multicolumn{6}{c}{Observables}    \\
\midrule
 & $\Delta a_{\mu}^{\rm SUSY}$ & $\Omega h^2$  & $\sigma_{\tilde{\chi}_1^0-n}^{\rm SI}$  & $\sigma_{\tilde{\chi}_1^0-n}^{\rm SD}$ & BR$(b\to s \gamma)$ \\
 \texttt{P1}~~~~~~~~ &  $16.3\times10^{-9}$  & 0.014     & $2.1\times10^{-12}$ pb  &  $5.0\times10^{-10}$ pb &  $3.4\times10^{-4}$\\
 \texttt{P2}~~~~~~~~ &  $17.7\times10^{-9}$  & 0.00034   & $1.7\times10^{-12}$ pb  &  $1.2\times10^{-8}$ pb &  $3.4\times10^{-4}$\\
\bottomrule
\end{tabularx}

\begin{tabularx}{\textwidth}{lcccccccc}
\toprule
\multicolumn{8}{c}{Ratio of event yields in signal regions of LHC searches to its 95\% observed limits }    \\
\midrule
 & $\sigma_{\tilde{\chi}_2^{0}\tilde{\chi}_1^{\pm}}$ & $\sigma_{\tilde{\chi}_1^{\pm}\tilde{\chi}_1^{\mp}}$  & $\sigma_{\tilde{\ell}\tilde{\ell}}$  & \texttt{SR-A08}  & \texttt{SR-A09} & \texttt{SR-SF-0J(160)} & \texttt{SR-SF-1J(160)} \\
 \texttt{P1}~&  0.399 pb  & 0.188 pb & 0.065 pb & 22.4 &  20.8 &  1.54 & 1.37 \\
 \texttt{P2}~ &  1.8$\times10^{-6}$ pb  & 7.53 pb & 0.0053 pb & $\ll0.1$ &  $\ll0.1$ &  0.43 & 0.31 \\
\bottomrule
\end{tabularx}
 \caption{Model parameters, mass spectrum and observables of benchmark points \texttt{P1} ($M_1=M_2=M_{1/2}$ at GUT scale) and \texttt{P2} ($M_1=5M_2=M_{1/2}$ at GUT scale) for $\tl{g}$SUGRA.}
  \label{Tab:gluinoSugra}
\end{table}

In Table ~\ref{Tab:gluinoSugra}, we list the model parameters at the GUT scale, the parameters at the SUSY scale, the masses, the observables of two benchmark points in the $\tl{g}$SUGRA scenario. Compared to the mSUGRA, the gluino mass $M_3$ is set above 3~\tev. As a result, the colored sparticles are heavy enough to avoid the constraints described in Sec.~\ref{sec:2}, including the LHC searches for gluino and stop, SM-like Higgs measurements and the BR$(B\to X_s\gamma)$ limits.

For benchmark point \texttt{P1}, we keep the relationship $M_1=M_2=M_{1/2}$ at the GUT scale, same as in the mSUGRA. Therefore, the LSP DM is bino-dominated and achieves the desired relic density via DM-slepton co-annihilation.  It is similar to the BL scenario of the pMSSM except for two differences that cause more stringent restrictions from the LHC direct searches. The first difference is that the lightest slepton is $\tilde{\tau}_1$ because of the universal slepton mass at the GUT scale. Thus, only the mass splitting between $\tilde{\tau}_1$ and $\tilde{\chi}_1^0$ is small, while the mass splittings between the first two generations of sleptons and $\tilde{\chi}_1^0$ are sizable, which leads to energetic lepton final states from slepton pair production at the LHC. In the last part of Table ~\ref{Tab:gluinoSugra}, we present ratios of event yields in the signal regions of the LHC searches to its 95\% observed limits. \texttt{SR-SF-0J(160)} and \texttt{SR-SF-1J(160)} are inclusive signal regions of $m_{T2}>160$ GeV in the search of electroweak production of sleptons decaying into final states with $2\ell + E_T^{\rm miss}$ with 139 fb$^{-1}$ LHC data~\cite{Aad:2019vnb}. The event yields in these two signal regions are calculated using the simulated events of slepton pair production 
\footnote{The $\tilde{\chi}_1^{\pm}\tilde{\chi}_1^{\mp}$ production can also generate final states with $2\ell + E_T^{\rm miss}$ and contribute to \texttt{SR-SF-0J(160)} and \texttt{SR-SF-1J(160)}. To illustrate that \texttt{P1} will still be excluded even if the wino mass is set to $>1$~\tev, we do not add $\tilde{\chi}_1^{\pm}\tilde{\chi}_1^{\mp}$ production to the slepton pair production for these two signal regions.}.
We can see that the ratios are larger than one, which means that this point is excluded by the slepton searches at the LHC. 
The second difference is that the masses of wino-like $\tilde{\chi}_1^{\pm}$ and $\tilde{\chi}_2^{0}$ are about 2 times of $m_{\tilde{\chi}_1^0}$. As demonstrated in the BL scenario of the pMSSM, $m_{\tilde{\chi}_1^0}$ needs to smaller than 400~\gev to explain the muon $g-2$ anomaly, which leads to $m_{\tilde{\chi}_1^{\pm}}\simeq m_{\tilde{\chi}_2^{0}} \simeq 2m_{\tilde{\chi}_1^{0}} < 800~\gev$ if $M_1=M_2$ at the GUT scale. In the latest searches for electroweak production of charginos and neutralinos~\cite{CMS-PAS-SUS-19-012}, the wino-like charginos with mass up to 1.4~\tev have been excluded if $\tilde{\chi}_1^{\pm}$ and $\tilde{\chi}_2^{0}$ dominantly (100\%) decay to the first two generations of sleptons. Despite $\tilde{\chi}_1^{\pm}$ and $\tilde{\chi}_2^{0}$ can decay to stau with about 30\% branching ratio, such a limit can only be relieved to about 1~\tev, much higher than the wino mass in this scenario. The \texttt{SR-A08} and \texttt{SR-A09} in Table~\ref{Tab:gluinoSugra} are signal regions of the CMS search for electroweak production of charginos and neutralinos in multilepton final states with 35.9 fb$^{-1}$ data~\cite{Sirunyan:2017lae}. The event yields of \texttt{P1} in these regions are 20 times higher the 95\% CL upper exclusion limits, because of the light winos.

As there is no strong motivation to set $M_1=M_2$ at the GUT scale, we can alter $M_2$ to avoid the LHC direct search constraints. A natural thought is to lift the wino mass $M_2$ beyond the LHC reaches. However, it does not help to avoid the constraints of slepton searches or even may make them worse, because the heavy winos will increase mass splittings between left- and right-handed sleptons, which makes sleptons easier to be found at the LHC. An alternative way is to decrease $M_2$ to compress the mass spectrum, such as \texttt{P2} in Table~\ref{Tab:gluinoSugra}. We set $M_1=5M_2$ at the GUT scale, and therefore the LSP DM is wino-dominated. Indirect detection constraints on gamma rays have set strong constraints on the wino-like DM. Nevertheless, here the LSP is only a tiny component of dark matter, which can escape such constraints~\cite{Fan:2013faa,Cohen:2013ama}. The constraints on sleptons are relaxed because sleptons decay to $\tilde{\chi}_1^{\pm}$ and $\tilde{\chi}_2^{0}$ instead of directly to $\tilde{\chi}_1^{0}$. Meanwhile, $\tilde{\chi}_1^{\pm}$ mainly decays to $\tilde{\chi}_1^{0} ud$, and $\tilde{\chi}_2^{0}$ mainly decays to $\tilde{\tau}_1^{\pm}\tau^{\mp}$. The long and complex decay chains make the final states of slepton pair production rather soft, and the event yields in \texttt{SR-SF-0J(160)} and \texttt{SR-SF-1J(160)} can be below their 95\% exclusion limits. So, in general, the muon $g-2$ anomaly can be explained in the $\tl{g}$SUGRA without conflicting with other constraints.

\item
Another type of extensions of the CMSSM are assuming non-universal soft SUSY breaking Higgs masses (NUHM) \cite{NUHM1,NUHM2}. In these extensions, relatively rare phenomena in the CMSSM parameter space can become much more 'mainstream' \cite{baer}. In the NUHM1 case, the effective Higgs masses are assumed to be universal at the GUT scale (though its value may  be different from the universal $m_0$) and thus there is one additional input parameter. While in the NUHM2 case, the Higgs scalar masses are independently non-universal, adding two additional input parameters. In both NUHM1 and NUHM2 scenarios, although the electroweak fine-tuning can possibly be small, the constraints from the gluino mass are still very stringent with a universal gaugino mass input at the GUT scale, just as in the CMSSM. The SUSY contribution to the muon $g-2$ is suppressed so it cannot account for the muon $g-2$ discrepancy.

\item 
The Focus Point (FP) SUSY, in which all squarks and sleptons may be multi-TeV without increasing the fine-tuning at the weak scale with respect to variation in the fundamental SUSY-breaking parameters \cite{FPGM1,FPGM2}, is originally proposed in the framework of gravity mediation. With specific types of GUT-scale boundary conditions, the RGE evolution drives $m^2_{H_u}$ to a value around $m_Z^2$ at the weak scale, almost independent of its initial GUT-scale value, implying that the weak scale in FP SUSY theories is not fine-tuned.  However, multi-TeV sleptons can hardly give large SUSY contributions to account for $\Delta a_\mu$ in the FP region of CMSSM. It is noted that the FP SUSY requires neither gaugino mass nor A-parameter unification, and also does not constrain the scalar masses that are only weakly coupled to the Higgs sector, such as the first and second generation squark and slepton masses \cite{Feng:1112.3021}. In scenarios with slight modification of FP region of CMSSM, in which all scalars have GUT-scale mass $m_0$ except for the smuons and muon sneutrino, it may resolve the muon $g-2$ discrepancy and at the same time be consistent with all current constraints, although the electroweakinos are still not very light due to the preserved gaugino ratio. Of course, the split SUSY, which set all the sfermions to be very heavy, also cannot account for the muon $g-2$ discrepancy.


\item In the NMSSM, in addition to the various MSSM contributions to $\Delta a_\mu$, a positive contribution to $\Delta a_\mu$ from two-loop diagrams involving a closed fermion loop can dominate the negative contributions from the one-loop Higgs diagrams \cite{Ellwanger:2009dp}. One-loop and two-loop light Higgs contributions are of opposite signs and interfere destructively. The sum of both contributions from the light CP-odd Higgs scalar $a_1$ has a positive maximum for $M_{a_1}\sim 6$ GeV though fairly insensitive to $M_{a_1}$ in the range of $4\sim 10$ GeV \cite{g-2:ellwanger}. In the constrained $Z_3$-invariant NMSSM (CNMSSM) , it can hardly generate such a light CP-odd Higgs scalar $a_1$ with a universal scalar mass parameter for both Higgs and sfermions. Therefore, the CNMSSM cannot give the NMSSM-specific large contributions to $\Delta a_\mu$.
\end{itemize}

\section{Muon $g-2$ in GMSB/AMSB extensions}   
In the above we surveyed over the muon $g-2$ in the CMSSM/mSUGRA and its extensions, 
where the boundary conditions are imposed at the GUT scale and the SUSY breaking 
is mediated through gravity. 
Finally, we briefly discuss the muon $g-2$ in other popular SUSY breaking mechanisms, namely the gauge mediated SUSY breaking (GMSB) mechanism and the anomaly mediated SUSY breaking (AMSB) mechanism: 

\begin{itemize}
\item
In the minimal GMSB model, the soft SUSY breaking spectrum at the messenger scale are given by
\beqa
M_i=\f{\al_i}{4\pi}\f{F}{M}N_{\bf 5}~,~{m}^2_{\tl{f}}=2\sum\limits_{i=1}^2C_{\tl{f}}^i \(\f{\al_i}{4\pi}\)^2\(\f{F}{M}\)^2 N_{\bf 5}, 
\eeqa
with $C_{\tl{f}}^i$ being the corresponding quadratic Casimir invariants. 
The gaugino masses at the messenger scale still satisfy the ordinary gaugino relations
\beqa
\f{M_1}{g_1^2}=\f{M_2}{g_2^2}=\f{M_3}{g_3^2}~.
\eeqa
It is obvious that the gaugino ratio $M_1:M_2:M_3\approx 1:2:6$ is preserved at the weak scale. Given the lower mass bound 2.2 TeV for gluino from the LHC, the bino should be heavier than 370 GeV. The soft SUSY breaking mass-squared for the sleptons at the messenger scale are much smaller than the mass-squared for the squarks. Typical low energy soft SUSY breaking mass spectrum versus the messenger scale can be seen in Fig.3 of \cite{GMSB}. Besides, as noted there, the mass ratios plotted in Fig.3 (apart from $m_{\tl{t}} /M_1$ ) are fairly independent of $F/M$ and $\tan\beta$. So, there are still allowed parameter space for the explanation of $\Delta a_\mu$ in GMSB. The main problem of GMSB is that the 125 GeV Higgs mass can hardly be explained with low stop masses, given the trilinear coupling input $A_t=0$ at the messenger scale. Heavy stop masses of order 5 TeV in the case of small $A_t$ also indicate heavy slepton masses and electroweakino masses, making it difficult to explain the $\Delta a_\mu$ anomaly. So the minimal GMSB needs to be extended to generate a non-vanishing $A_t$ for TeV scale stop masses, relaxing the constraints on $\Delta a_\mu$ anomaly from the 125 GeV Higgs. A popular extension of minimal GMSB is to introduce additional Yukawa mediation contributions so that a non-vanishing trilinear coupling $A_t$ can be generated at the messenger scale \cite{Evans:2012hg,Evans:2011bea,Jelinski:2011xe,Kang:2012ra}.

\begin{figure}[t]
\label{fig-amsb} 
\centering 
\includegraphics[width=.49\textwidth]{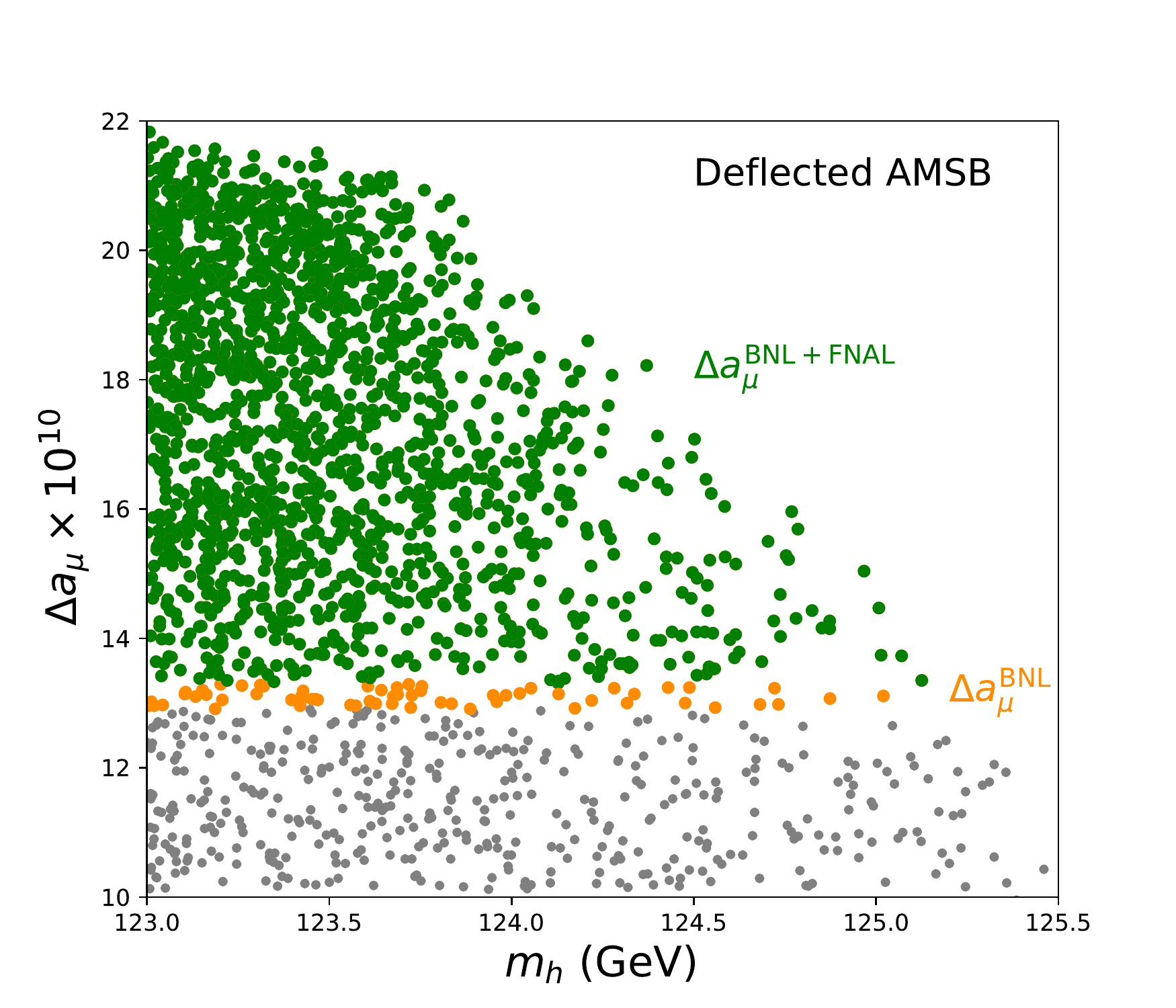}
\includegraphics[width=.49\textwidth]{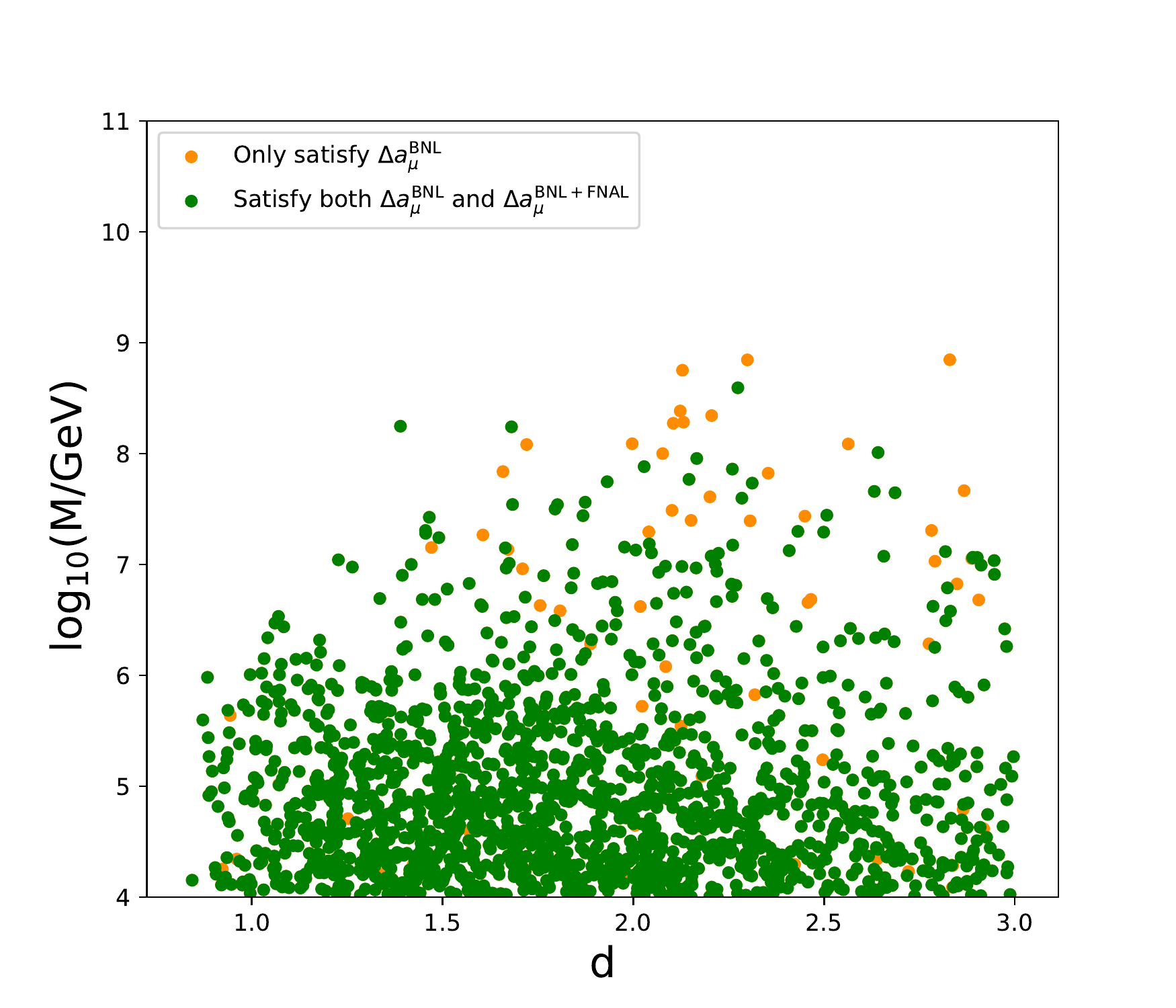}
\caption{\label{fig:AMSB} Scatter plots of the parameter space in the deflected AMSB \cite{Wang:2015nra}. All the points survive the collider and dark matter constraint.
The green points satisfy both BNL+FNAL and BNL muon $g-2$ results, while the orange points only satisfy the BNL result. }
\end{figure}

\item
Gravity can generate the soft SUSY breaking masses by the auxiliary field of the compensator multiplet. Such a $'pure'$ gravity mediation scenario with negligible contributions from direct non-renormalizable contact terms is called the anomaly mediated SUSY breaking (AMSB). Although very predictive, the pure anomaly mediation is unfortunately bothered by the tachyonic slepton problem ~\cite{tachyonslepton-1,tachyonslepton-2,tachyonslepton-3,tachyonslepton-4} and must be extended. Its non-trivial extensions with messenger sectors, namely the deflected AMSB~\cite{dAMSB-1,dAMSB-2,okada} can elegantly solve such a tachyonic slepton problem through the deflection of the RGE trajectory ~\cite{deflect:RGE-invariance}. The gaugino mass ratios at the messenger scale depend on both the deflection parameters and the number of messengers. Ordinary gaugino ratio bounds can be relaxed in the deflected AMSB. Besides, the mass-squared for the sleptons at the messenger scale are still much smaller than the mass-squared for the squarks. In the deflected AMSB, a large value of $A_t$ is automatic at the messenger scale. So the muon $g-2$ can get sizable contributions in the deflected AMSB \cite{Wang:2015nra,Wang:2016otm,Ning:2017dng}. 
In Fig. \ref{fig-amsb} we re-plot the results of the deflected AMSB \cite{Wang:2015nra}, showing the impact of the new measurement of the muon $g-2$. We see that this model
can readily explain the muon $g-2$ at $2\sigma$ CL. The results are obtained by scanning following parameter space,
\beqa
&3 \ge d \ge -3,~ N_F \ge 5,~ 10^4\gev\le M \le 10^{16}\gev , \nonumber \\ 
&10 \tev < F_{\phi} < 500 \tev,~  2 \le \tan\beta \le 40,~ {\rm sign}(\mu) = {\rm sign}(M_2), \nonumber
\eeqa
where $d$ indicates the deflection parameter, $N_F$ stands for the flavor number,  $M$ is the messenger scale, and $F_{\phi}$ is the auxiliary compensator field within the graviton supermultiplet. 
Here, the displayed samples satisfy both upper and lower limits on the DM relic density, the lower bounds of LEP on neutralino, charginos and sleptons masses, constraints of the precision electroweak measurements and the Higgs mass measurements. The latest searches for SUSY at LHC are not applied to all the samples, but we have checked that some of the samples can avoid such constraints because of compressed mass spectrum. The ranges of mass parameters for samples consistent with the new measurement of the muon $g-2$ are shown in Tab.\ref{Tab:AMSB}.

\begin{table}[t]
\centering
\begin{tabular}{ccccc}
\toprule
$m_h$ & $m_{H}$ & $m_{A}$ & $m_{H^{\pm}}$ & $m_{\tilde{g}}$\\
$[123.0,125.0]$ & $[1678.3,2650.2]$ & $[1678.3,2650.2]$ & $[1680.4,2651.6]$ & $[2815.9,4546.4]$ \\
\midrule
$m_{\tilde{\chi}_1^0}$ & $m_{\tilde{\chi}_2^0}$ & $m_{\tilde{\chi}_3^0}$ & $m_{\tilde{\chi}_4^0}$ & $m_{\tilde{\chi}_1^{\pm}}$ \\
$[59.7,119.3]$ & $[340.3,552.0]$ & $[1683.0,2679.2]$ & $[1684.7,2680.0]$ & $[340.3,552.0]$ \\
\midrule
$m_{\tilde{\ell}_1}$ & $m_{\tilde{\ell}_2}$ & $m_{\tilde{\tau}_1}$ & $m_{\tilde{\tau}_2}$ & $m_{\tilde{\nu}_\tau}$\\
$[137.9,226.9]$ & $[385.5,592.4]$ & $[100.0,133.0]$ & $[393.9,606.4]$ & $[376.6,582.6]$ \\
\midrule
$m_{\tilde{t}_1}$ & $m_{\tilde{t}_2}$ & $m_{\tilde{b}_1}$ & $m_{\tilde{b}_2}$ & $m_{\tilde{u}_1}$\\
$[2088.7,3386.5]$ & $[2362.6,3763.1]$ & $[2516.4,4044.1]$ & $[2546.9,4085.3]$ & $[2545.7,4084.6]$\\
\bottomrule
\end{tabular}
\caption{Ranges of mass for the samples in the  deflected AMSB consistent with the new measurement of the muon $g-2$, i.e. green points in Fig.\ref{fig-amsb}. All the masses are in \gev. }\label{Tab:AMSB}
\end{table}

\end{itemize}

\section{Conclusion}
In light of the E989 experimental result for the muon $g-2$  and  
the LHC results for the superparticle searches, the SM-like Higgs boson and Br$(b\to X_s\gamma)$,  we revisited various GUT-scale constrained SUSY models which assume universal boundary conditions at the GUT scale.
We first demonstrated the tension of the typical one, the so-called CMSSM/mSUGRA, confronting with these measurements.
After discussing the possible ways to alleviate such a tension and showing experimental constraints that are related directly to the SUSY contribution to the muon $g-2$,  including the DM relic density, the DM direct detections, the electroweakino and slepton searches and electroweak vacuum stability, we surveyed several extensions of the CMSSM/mSUGRA with different types of universal boundary conditions at the GUT scale, which can satisfy most or all of these experimental constraints. Finally,  
 we briefly discussed the muon $g-2$ in other popular SUSY breaking mechanisms, namely the GMSB and AMSB mechanisms and their extensions. 
 
\addcontentsline{toc}{section}{Acknowledgments}
\acknowledgments

We would like to thank Zhuang Li for his helping on finding $\tl{g}$SUGRA benchmark points. This work was supported by the National Natural Science Foundation of China 
(NNSFC) under grant Nos. 11675147,  11821505, 12075300 and 12075213,  
by the Key Research Project of Henan Education Department for colleges and universities under grant number 21A140025,
by Peng-Huan-Wu Theoretical Physics Innovation Center (12047503),
by the CAS Center for Excellence in Particle Physics (CCEPP), 
by the CAS Key Research Program of Frontier Sciences, 
and by a Key R\&D Program of Ministry of Science and Technology of China
under number 2017YFA0402204.

\addcontentsline{toc}{section}{References}
\bibliographystyle{JHEP}
\bibliography{bibliography}

\end{document}